\documentclass[
 reprint,
superscriptaddress,
 nofootinbib,
 amsmath,amssymb,
 aps,
pra,
]{revtex4-2}

\usepackage{graphicx}
\usepackage{dcolumn}
\usepackage{bm}
\usepackage{xcolor}
\usepackage{braket}
\usepackage{amsmath, amssymb,mathcomp,enumerate}
\usepackage{bbold}
\usepackage{soul}
\usepackage{siunitx}
\usepackage[colorlinks=true,allcolors=black]{hyperref}
\usepackage{comment}
\usepackage{array}
\usepackage[version=4]{mhchem}
\usepackage{booktabs}
\usepackage{makecell}
\usepackage{svg}
\usepackage[T1]{fontenc}

\usepackage[symbol*]{footmisc}
\setfnsymbol{wiley}

\DeclareSIUnit\bar{bar}

\newcommand{\Ct}{C_{t}}
\newcommand{\Cs}{C_{s}}
\newcommand{\ECq}{E_{Cq}}
\newcommand{\ECa}{E_{Ca}}
\newcommand{\EJ}{E_{J}}
\newcommand{\La}{L_{a}}
\newcommand{\LJ}{L_{J}}

\newcommand{\bphq}{\hat{\varphi}_q}
\newcommand{\bpha}{\hat{\varphi}_a}

\newcommand{\bnq}{\hat{n}_q}
\newcommand{\bna}{\hat{n}_a}
\newcommand{\wc}{\omega_c}
\newcommand{\wq}{\omega_q}
\newcommand{\wa}{\omega_a}
\newcommand{\ga}{g_{ac}}
\newcommand{\proba}{\text{Prob}}

\begin{document}

\title{High-power readout of a transmon qubit using a nonlinear coupling}

\author{Cyril Mori}
\altaffiliation{These authors contributed equally to this work}
\affiliation{Univ. Grenoble Alpes, CNRS, Grenoble INP, Institut N\'eel, 38000 Grenoble, France}

\author{Vladimir Milchakov}
\altaffiliation{These authors contributed equally to this work}
\affiliation{Univ. Grenoble Alpes, CNRS, Grenoble INP, Institut N\'eel, 38000 Grenoble, France}
\author{Francesca D'Esposito}
\affiliation{Univ. Grenoble Alpes, CNRS, Grenoble INP, Institut N\'eel, 38000 Grenoble, France}
\author{Lucas Ruela}
\affiliation{Univ. Grenoble Alpes, CNRS, Grenoble INP, Institut N\'eel, 38000 Grenoble, France}
\author{Shelender Kumar}
\affiliation{Univ. Grenoble Alpes, CNRS, Grenoble INP, Institut N\'eel, 38000 Grenoble, France}
\author{Vishnu Narayanan Suresh}
\affiliation{Univ. Grenoble Alpes, CNRS, Grenoble INP, Institut N\'eel, 38000 Grenoble, France}
\author{Waël Ardati}
\affiliation{Univ. Grenoble Alpes, CNRS, Grenoble INP, Institut N\'eel, 38000 Grenoble, France}
\author{Dorian Nicolas}
\affiliation{Univ. Grenoble Alpes, CNRS, Grenoble INP, Institut N\'eel, 38000 Grenoble, France}
\author{Giulio Cappelli}
\affiliation{Univ. Grenoble Alpes, CNRS, Grenoble INP, Institut N\'eel, 38000 Grenoble, France}
\author{Arpit Ranadive}
\altaffiliation{Present address: Google Quantum AI, Goleta, California 93117, USA.}
\affiliation{Univ. Grenoble Alpes, CNRS, Grenoble INP, Institut N\'eel, 38000 Grenoble, France}
\author{Gwenael Le Gal}
\affiliation{Univ. Grenoble Alpes, CNRS, Grenoble INP, Institut N\'eel, 38000 Grenoble, France}
\author{Martina Esposito}
\altaffiliation{Present address: CNR-SPIN Complesso di Monte S. Angelo, 80126 Napoli, Italy.}
\affiliation{Univ. Grenoble Alpes, CNRS, Grenoble INP, Institut N\'eel, 38000 Grenoble, France}
\author{Quentin Ficheux}
\affiliation{Univ. Grenoble Alpes, CNRS, Grenoble INP, Institut N\'eel, 38000 Grenoble, France}
\author{Nicolas Roch}
\affiliation{Univ. Grenoble Alpes, CNRS, Grenoble INP, Institut N\'eel, 38000 Grenoble, France}
\author{Tom\'as Ramos}
\affiliation{Institute of Fundamental Physics IFF-CSIC, 28006 Madrid, Spain}
\author{Olivier Buisson}
\email{Contact author: olivier.buisson@neel.cnrs.fr}
\affiliation{Univ. Grenoble Alpes, CNRS, Grenoble INP, Institut N\'eel, 38000 Grenoble, France}

\date{\today}

\begin{abstract}
The field of superconducting qubits is constantly evolving with new circuit designs. However, when it comes to qubit readout, the use of simple transverse linear coupling remains overwhelmingly prevalent. This standard readout scheme has significant drawbacks: in addition to the Purcell effect, it suffers from a limitation on the maximal number of photons in the readout mode, which restricts the signal-to-noise ratio (SNR) and the Quantum Non-Demolition (QND) nature of the readout. Here, we explore the high-power regime by engineering a nonlinear coupling between a transmon qubit and its readout mode. Our approach builds upon previous work by Dassonneville et al. [Physical Review X 10, 011045 (2020)], on qubit readout with a non-perturbative cross-Kerr coupling in a transmon molecule. We demonstrate a readout fidelity of 99.21\% with 89 photons utilizing a parametric amplifier. At this elevated photon number, the QND nature remains high at 96.7\%. Even with up to 300 photons, the QNDness is only reduced by a few percent. This is qualitatively explained by deriving a critical number of photons associated with the nonlinear coupling, yielding a theoretical value of $\bar{n}_r^\text{crit} = 377$ photons for our sample's parameters. These results highlight the promising performance of the transmon molecule in the high-power regime, establishing it as a compelling platform for high-fidelity qubit readout.
\end{abstract}

\maketitle

\section{Introduction}


For any quantum algorithm, readout errors directly impact the computation's result. On the one hand, readout fidelity limits the performance of any active error correction code \cite{Shor1995,Gottesman1997}.
On the other hand, quantum non-demolition (QND) errors impact the performance of qubit initialization with pre-selection \cite{Johnson2012} or active feedback \cite{Riste2012}. Therefore, achieving a high-fidelity but also high QNDness readout is key to improving performance in quantum computing.

In superconducting qubits, the most common readout scheme comprises a transmon coupled to a microwave resonator through a transverse linear coupling \cite{Koch2007}. For a large qubit-resonator detuning, this scheme is QND under a first-order perturbative approximation. The dispersive shift induced by the transmon qubit on the resonator's frequency is given by $\chi=g^2 \alpha_q/[\Delta (\Delta+\alpha_q)]$, where $g$ is the coupling strength, $\Delta$ the qubit-resonator detuning, and $\alpha_q$ the transmon's anharmonicity. However, a large detuning $\Delta\gg g$ leads to a reduced dispersive shift $\chi\ll g$. Furthermore, the transverse linear coupling dresses the bare qubit state with the resonator states. This induces bit-flip errors due to Purcell effect. In addition, the approximation is only valid when the readout microwave power, defined by the average number of readout photons $\bar{n}_r$, is lower than $\bar{n}_\text{std}^\text{crit}=\Delta^2/(2g)^2$. This critical number of photons $\bar{n}_\text{std}^\text{crit}$ limits the readout power and SNR (signal-to-noise ratio), as observed experimentally \cite{Swiadek2024}. The usual approach so far has been to work in the low power regime to preserve QNDness and compensate the SNR by adjusting the circuit parameters.


However, the state-of-the-art in readout fidelity is evolving towards higher readout powers and several strategies have been implemented to attain high-power readout. Ref. \cite{Kurilovich2025,Connolly2025} showed that working in a regime of extremely high detuning can protect against measurement-induced state transitions (MIST). Using a detuning $\Delta/\omega_q \approx 11$, with $\omega_q$ the qubit's frequency, they achieved a readout fidelity of 99.93\% with 92 photons. Alternatively, in Ref. \cite{Gusenkova2021,Takmakov2021}, a transversely coupled fluxonium displays a readout fidelity of 96\% using 74 photons, without any parametric amplification.

In parallel, several alternative transmon readout schemes have been proposed and implemented. They rely on nonlinear couplings instead of simple transverse coupling \cite{Diniz2013,Dassonneville2020,Ye2024,Pfeiffer2024,Salunkhe2025}. Such schemes present different physics and benefits compared to the standard transmon readout scheme. In particular, in Ref. \cite{Salunkhe2025}, a fidelity of 98.3\% is attained without parametric amplifier and with a readout power of 30 photons, indicating that their readout scheme is robust to high readout powers. In the current work, a transmon is measured through a purely nonlinear coupling named the cos$\varphi$-coupling, implemented using a transmon molecule circuit \cite{Dassonneville2020,Dassonneville2023}. The readout scheme presented in this article is robust to high numbers of readout photons. This observation is qualitatively explained by deriving the critical number of photons associated to the $\cos\varphi$-coupling, yielding a value in the hundreds of photons.



The $\cos\varphi$-coupling has been studied theoretically as a promising candidate for high-fidelity qubit readout \cite{Dassonneville2020,Diniz2013,Didier2015}. It offers Purcell protection due to its purely nonlinear properties and is predicted to present better QNDness than transverse coupling at high readout powers \cite{Chapple2024}. Interestingly, the critical number of photons introduced in the transverse-coupling readout scheme \cite{Blais2004,Gambetta_2006} is not a limit anymore for the $\cos\varphi$-coupling since no perturbation theory is applied. 

A first implementation of the transmon molecule readout scheme has been demonstrated in \cite{Dassonneville2020,Dassonneville2023}, showcasing the QNDness of the $\cos\varphi$-coupling in the low power regime. In this article, we present a new generation of devices optimized to operate in the high-power readout regime with high fidelity and high QNDness. We report a readout fidelity of 99.21\%, measured at a power of 89 photons and with parametric amplifier. The QNDness, at such high power, is estimated at 96.7\%. Furthermore, a study of the QNDness as a function of the readout power reveals only few percents of errors up to 300 readout photons. This indicates that the transmon molecule readout scheme is a promising candidate for quantum computing architectures. This robustness to readout photons can be explained by analyzing the theoretical limitation of the $\cos\varphi$-coupling as a function of the number of photons. A novel critical number of readout photons is computed as a function of the circuit's parameters $\bar{n}_r^\text{crit} = 377$. It is associated to the validity of the Taylor development yielding the cross-Kerr coupling term used for dispersive readout. This value is more than one order of magnitude higher than the usual critical number of photons for most transverse readout schemes.

\begin{table}[t]
\centering
\begin{tabular}{c@{\hspace{2em}}c} 
\toprule
\toprule
Parameter name & Value \\
\toprule
$\wq/(2\pi)$ & \SI{2.0332}{\giga\hertz} \\
$\ECq/(2\pi\hbar)$ & \SI{0.0731}{\giga\hertz} \\
$E_{Jq}/(2\pi\hbar)$ & \SI{7.68}{\giga\hertz} \\
$\omega_r/(2\pi)$ & \SI{7.290}{\giga\hertz} \\
$\alpha_r/(2\pi)$ & \SI{-0.00682}{\mega\hertz} \\
$\kappa_r/(2\pi)$ & \SI{17.9}{\mega\hertz} \\
$\chi_{qr}/(2\pi)$ & \SI{-0.77}{\mega\hertz}\\
$\bar{n}^\text{crit}_r$ & 377 photons \\
$F_\text{ps}$ & 99.21\% \\
\bottomrule
\bottomrule
\end{tabular}
\caption{Key sample parameters.}
\label{tab:main_params}
\end{table}

\section{The \texorpdfstring{cos$\varphi$}{cosϕ}-coupling}
\label{sec:theory}

\begin{figure}[t]
\centering
\includegraphics[width=1\columnwidth]{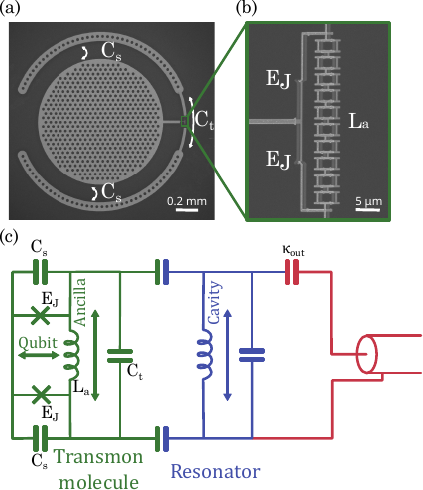}
\caption{\label{fig:circuit} (a) Scanning electron microscope image of a transmon molecule sample. (b) Zoom on the loop of the transmon molecule: the inductor (chain of SQUIDs) and the two Josephson junctions. (c) Lumped elements circuit of the full transmon molecule readout scheme. The input coupling $\kappa_\text{in}$ is not represented since $\kappa_{in}\ll\kappa_\text{out}$.}
\end{figure}

The transmon molecule is a bimodal circuit which consists of two transmons coupled through an LC circuit, as shown in Fig. \ref{fig:circuit}(c). Since there is a coupling between two transmon ``artificial atoms'', the circuit is named transmon molecule. The two nominally identical transmons have capacitances $\Cs$ and Josephson energies $\EJ$. The LC oscillator has a capacitance $\Ct$ and an inductance $\La$. In this paper, we consider that the circuit's superconducting loop is biased at integer-flux. The regime of non-zero flux bias is treated in detail and explored in previous works \cite{Dassonneville2020,Dumur2016}. At zero-flux, the full circuit Hamiltonian has been derived in \cite{Diniz2013,Dassonneville2020} and can be written as:
\begin{align}
\label{eq:ham_circuit}
\begin{split}
    \hat{H}_{\text{tm}} &= 4\ECq\bnq^2 - E_{Jq}\cos(\bphq)\\
    & + 4\ECa\bna^2 - 2\EJ\left[\cos(\bpha)-\frac{\LJ}{\La}\bpha^2\right]\\
    & - 2\EJ\left[\cos(\bphq)-1\right] \left[\cos(\bpha)-1\right].
\end{split}
\end{align}
The first two lines describe the two circuit modes, called qubit mode (noted $q$) and ancilla mode (noted $a$), respectively. The modes are defined by their number operators $\bnq$, $\bna$ and their phase operators $\bphq$, $\bpha$. Their charging energies are $\ECq=e^2/(4\Cs)$ and $\ECa=e^2/(8\Ct+4\Cs)$. The qubit mode has an effective Josephson energy $E_{Jq}=2\EJ$. The Josephson junction inductance is defined as $L_J=\varphi_0^2/\EJ$ with the reduced flux quantum $\varphi_0=\hbar/(2e)$. The qubit mode has exactly the Hamiltonian of a transmon. In contrast, the ancilla mode has an additional potential term coming from the inductance. Since $L_J/\La\gg 1$, this mode is very linear, so it will be used to perform readout. The sample's main parameters are shown in Table \ref{tab:main_params}.

The two modes are coupled through a cos$\varphi$-coupling term, which is purely nonlinear. As shown in Appendix \ref{apx:theory}.1, it leads to a non-perturbative cross-Kerr coupling of the form $2\chi_{qa} \hat{q}^\dag\hat{q} \hat{a}^\dag\hat{a}$, where $\hat{q}$ (resp. $\hat{a}$) is the qubit (resp. ancilla) mode's annihilation operator in the limit of small quantum fluctuations. Importantly, the coupling's strength $\chi_{qa}=- \sqrt{\ECq\ECa/(1+2\LJ/\La)}/\hbar$ is independent of the detuning between the modes and is tunable by adjusting the circuit parameters. 

A third mode is added to the system: a resonator with strong transverse coupling to the ancilla, and decoupled from the qubit by symmetry. This results in an absence of Purcell effect on the qubit. Due to their strong coupling, the ancilla mode and resonator hybridize, as described in Appendix \ref{apx:theory}.2. The resulting two modes inherit $\cos\varphi$-coupling to the qubit as well as a coupling to the output line. This means they can both act as readout modes. In our experiment, we choose the most linear of the hybrid modes as our readout mode. By considering that the other one remains in the vaccuum state, we get the following reduced Hamiltonian:
\begin{align}
\label{eq:ham_twomode}
\begin{split}
    \frac{\hat{H}_{\text{reduced}}}{\hbar} &= \wq \hat{q}^\dagger \hat{q} + \frac{\alpha_q} {2} \hat{q}^\dagger\hat{q}^\dagger \hat{q}\hat{q} \\
    &+\omega_r\hat{c}_r^\dag\hat{c}_r +  \frac{\alpha_r} {2} \hat{c}_r^\dagger\hat{c}_r^\dagger \hat{c}_r\hat{c}_r \\
    &+ 2\chi_{qr} \hat{q}^\dagger \hat{q} \hat{c}_r^\dag\hat{c}_r,
\end{split}
\end{align}
where $\wq$ is the qubit's resonant frequency, $\alpha_q$ is its transmon anharmonicity, $\omega_r$ is the readout mode's resonant frequency, $\alpha_r$ is the readout mode's anharmonicity, $\hat{c}_r$ is its annihilation operator and $\chi_{qr}$ is the strength of the cross-Kerr coupling. Eq. (\ref{eq:ham_twomode}) is exactly the same Hamiltonian as for a qubit dispersively coupled to a readout resonator \cite{Wallraff2005}, but it does not rely on any high-detuning approximation. Furthermore, the cross-Kerr term is not perturbative, since it comes from the $\cos\varphi$-coupling which is itself purely nonlinear. For the chosen readout mode, the effective anharmonicity is $\alpha_r = \alpha_a \sin^4\theta$, with $\alpha_a$ the ancilla mode's anharmonicity, and its effective cross-Kerr coupling strength is $\chi_{qr}=\chi_{qa}\sin^2(\theta)$. Both of these quantities depend on the hybridization mixing angle $\theta=(1/2)\arctan[2g_{ac}/(\omega_c-\omega_a)]$, where $\omega_c$ (resp. $\omega_a$) is the bare cavity (resp. ancilla) frequency and $g_{ac}$ is the strength of the cavity-to-ancilla coupling.


\section{Implementation}
\label{sec:implementation}


The transmon molecule circuit is fabricated using aluminum on silicon. The sample is shown in Fig. \ref{fig:circuit}(a). The two Josephson junctions are $\ce{Al}/\ce{AlO}_x/\ce{Al}$ junctions made using Dolan bridge technique \cite{Niemeyer1976,Dolan1977}. The inductor is made from a chain of SQUIDs using bridge-free technique \cite{Lecocq2011}, as shown in Fig. \ref{fig:circuit}(b). The complete fabrication process is described in Appendix \ref{apx:fab}.

The transmon molecule is coupled to a 3D microwave resonator. The 3D resonator is a copper rectangular microwave cavity, which can be probed in transmission. The pins of the cavity's two ports have been adjusted to achieve the targeted input $\kappa_\text{in}/(2\pi)=\SI{0.153}{\mega\hertz}$ and output coupling $\kappa_\text{out}/(2\pi)=\SI{13.0}{\mega\hertz}$, with $\kappa_\text{out}\gg\kappa_\text{in}$. A coil made of thin superconducting wire is wrapped around the cavity for flux biasing the sample. The cavity's detailed characteristics and the full microwave setup are given in Appendix \ref{apx:setup}.

The transmon molecule is carefully aligned inside the 3D cavity such that the ancilla mode is strongly coupled to the resonator whereas the qubit mode is completely decoupled from the resonator, thanks to the orthogonality of the dipole moments of the qubit and ancilla modes [see Fig. \ref{fig:circuit}(c)].


The previous transmon molecule implementation in Ref. \cite{Dassonneville2020} was limited by the qubit's lifetime of $T_1=\SI{3.3}{\micro\second}$. This work's sample was optimized to go beyond this limitation. The sample's design presents a circular shape to minimize any remnants of direct transverse coupling between qubit and cavity (Appendix \ref{apx:optim}). The fabrication was optimized to reduce dielectric and TLS-induced losses (Appendix \ref{apx:fab}). The sample's electrical parameters were tuned to make the readout mode more linear and to minimize the effect of residual Purcell decay (Appendix \ref{apx:optim}). This optimized transmon molecule presents a relaxation time $T_1=\SI{124.5}{\micro\second}$ (Appendix \ref{apx:coherence}).

\section{Readout performance}
\label{sec:readout}


\subsection{Readout fidelity}

\begin{figure}[t]
\centering
\includegraphics[width=1\columnwidth]{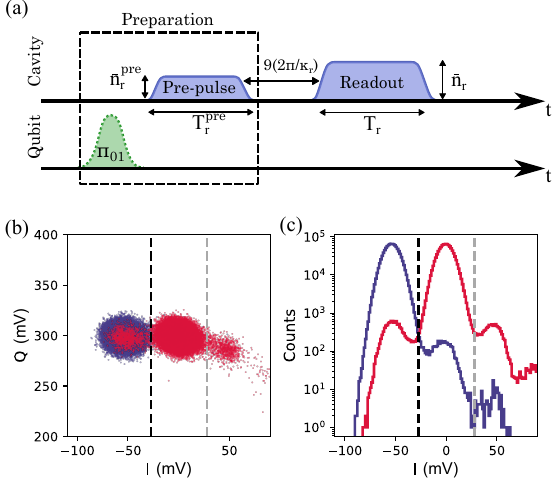}
\caption{\label{fig:fid} (a) Pulse sequence for the readout fidelity measurement. The state is prepared with an optional $\pi$-pulse (used when preparing $\ket{1}$) followed by a readout pulse used for pre-selecting the qubit state. (b) IQ plane representation of the single-shot readout fidelity measurement. The indigo and red points represent the prepared $\ket{0}$ and $\ket{1}$ state, respectively. The black dashed line is the binary threshold used to discriminate between $\ket{0}$ and $\ket{1}$. The grey dashed line is an additional threshold used to separate leakage states from the computational space. (c) Histogram representation of the single-shot readout fidelity measurement. The experiment was repeated $10^6$ times with a prepared $\ket{0}$-state and $10^6$ for a prepared $\ket{1}$ state. }
\end{figure}

The qubit state is prepared with an optional $\pi_{01}$-pulse followed by a pre-selection pulse, as illustrated in Fig. \ref{fig:fid}(a). This preparation is followed by a waiting time of $\SI{500}{\nano\second}\approx9 (2\pi/\kappa_r)$ to deplete the readout mode before the readout pulse. To improve the SNR, we use a traveling-wave parametric amplifier (TWPA) based on ``reverse-Kerr'' phase-matching \cite{Ranadive2022}, as shown in the setup (see Appendix \ref{apx:setup}).

The measurement results are shown in Fig. \ref{fig:fid}(b) and Fig. \ref{fig:fid}(c). The readout pulse's parameters are optimized following the procedure described in Appendix \ref{apx:ro_optim}. By setting a linear threshold between the $\ket{0}$ and $\ket{1}$ gaussians, the data points are classified into either of the two states. Then we can compute the standard binary readout fidelity $F=1-[\proba(1|0)+\proba(0|1)]/2$, where $\proba(i|j)$ is the conditional probability of measuring $\ket{i}$, having prepared $\ket{j}$. A fidelity of $F=99.21\pm0.01$\% was measured using an optimized square readout pulse of duration $T_r=\SI{400}{\nano\second}$ and power $\bar{n}_r=89$ photons (see Appendix \ref{apx:ro_optim} for more details on the optimization of the fidelity). The readout power is expressed in photon number by using the AC-Stark shift calibration described in Appendix \ref{apx:nphcalib}.

It is also worth noting that the maximum fidelity is measured using 89 photons, which is at least one order of magnitude higher than most state-of-the-art high-fidelity measurements relying on standard dispersive coupling \cite{Walter2017,Spring2024,Swiadek2024}. This power corresponds to an AC Stark shift of \SI{-138}{\mega\hertz} on the qubit's resonant frequency. Even though this shift is significantly larger than the qubit's anharmonicity $\alpha_q/(2\pi)=\SI{-73.1}{\mega\hertz}$, the readout fidelity is not noticeably affected. 

The readout fidelity computation can be refined by taking into account leakage errors. These errors can be seen in Fig. \ref{fig:fid}(c), as additional smaller gaussians besides $\ket{0}$ and $\ket{1}$, appearing on the right side of the IQ plane. Those gaussians correspond to leakage towards higher transmon states outside the computational space, which were identified by preparing and measuring the transmon in states $\ket{2}$ and above. To account for such errors, an additional threshold is set in the IQ plane to differentiate between the computational space $\{\ket{0};\ket{1}\}$ and all the leakage states $\ket{l}$. All probabilities are thus re-computed to account for three ``states'' $\ket{0}$, $\ket{1}$, $\ket{l}$ and are noted $\proba_{3s}$. The leakage probabilities are measured as $\proba_{3s}(l|0)=0.01\%$ and $\proba_{3s}(l|1)=1.0\%$. 
In practice, these types of errors can be simply post-selected out, as is done in Ref. \cite{Krinner2022,Miao2023}. In this case, the fidelity definition becomes $F_\text{ps}=[\proba_{3s}(1|1)+\proba_{3s}(0|0)]/[2-\proba_{3s}(l|0)-\proba_{3s}(l|1)]$, after normalizing by the total number of measurements which stayed in the computational space. The resulting fidelity is $F_\text{ps}=99.21\pm0.01$\%. Up to the significant digits, this is identical to $F$ because the correction $\proba_{3s}(1|1)=\proba(1|1)-\proba_{3s}(l|1)$ in the numerator is almost exactly compensated by the renormalization in the denominator.

\subsection{QNDness measurements}

\begin{figure}[t]
\centering
\includegraphics[width=1\columnwidth]{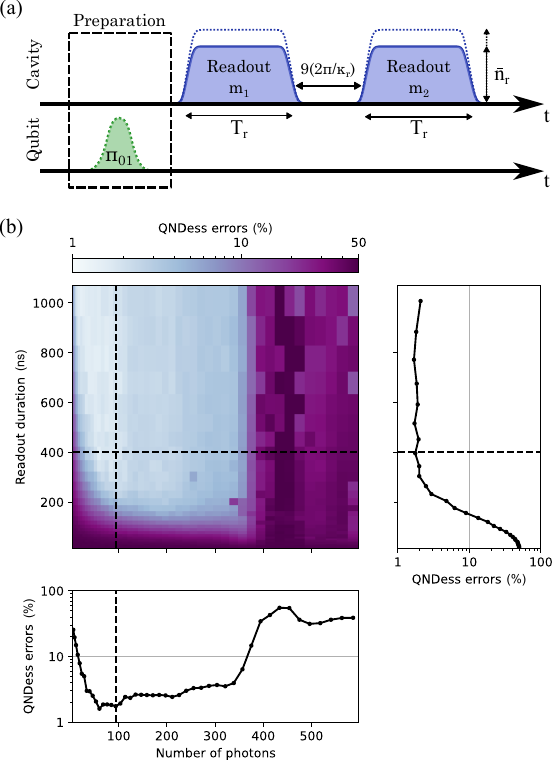}
\caption{\label{fig:qnd_vs_params} (a) Pulse sequence for the QNDness measurement. The state is prepared with an optional $\pi$-pulse (used to prepare $\ket{1}$) and then two readouts are performed in a row. The readout pulses have identical duration $T_r$ and power $\bar{n}_r$. (b) Colorplot of QNDness errors as a function of the readout pulses' length $T_r$ and power $\bar{n}_r$. The readout parameters showing maximal readout fidelity are indicated with black dashed lines. On the right, a slice of the QNDness errors is plotted as function of readout duration, at the optimal readout power $\bar{n}_r=89$ photons. On the bottom, a slice of the QNDness errors is plotted as function of readout power, at the optimal readout duration $T_r=\SI{400}{\nano\second}$. }
\end{figure}

To complete the characterization of the readout performance, we also measure the sample's QNDness. There are several different protocols to characterize the readout's QNDness \cite{Pereira2022,Pereira2023,Hazra2024} and the one used here is from Ref. \cite{Touzard2019,Dassonneville2020}. As shown in Fig. \ref{fig:qnd_vs_params}(a), the qubit is initialized using an optional $\pi$-pulse without pre-selection, then two identical readout pulses are sent in a row. The initialization here does not need to be precise since it is only meant for evening out the statistics between $\ket{0}$ and $\ket{1}$-state measurements. If we note the measurement result $m_1$ (resp. $m_2$) for the first (resp. second) readout pulse, then we can compute the probabilities $\proba(m_1,m_2)$. The QNDness is quantified by the probability of having correlated results on both readout pulses, which we will note $P_\text{QND}=\proba(0,0)+\proba(1,1)$ \cite{Touzard2019,Dassonneville2020,Pereira2022}. Note that the statistics measured with and without $\pi$-pulse are merged for clarity. This definition is equivalent to that of Ref. \cite{Touzard2019}, where a factor $1/2$ appears due to the $\proba(0,0)$ and $\proba(1,1)$ being computed from separate statistics. With this definition,  we get a binary QNDness $P_\text{QND}=98.2\pm0.2$\%.

Once again, this definition can be extended to account for leakage. Thresholds are added to discriminate between the computational space and the leakage states for $m_1$ and for $m_2$. All the probabilities are re-computed and noted $\proba_{3s}(m_1,m_2)$. Furthermore, the qubit starts from a thermal state so it can be in a leakage state even before the readout pulse $m_1$. Since the QNDness should only quantify the effect of the readout on computational states $\ket{0}$ and $\ket{1}$, the data points where $m_1=l$ are post-selected out. The points where $m_2=l$ are kept in the statistics and counted as errors. The resulting QNDness is $P^\text{ps}_\text{QND}=[\proba_{3s}(0,0)+\proba_{3s}(1,1)]/[1-\proba_{3s}(l,0)-\proba_{3s}(l,1)-\proba_{3s}(l,l)]=96.7\pm0.2$\%. In both cases, our QNDness is above 96\%, even though the readout power is at 89 photons. At such populations in the readout mode, one could expect instead that the QNDness would plummet due to exchanges of energy causing qubit decoherence \cite{Walter2017,Swiadek2024}.

To further understand how the readout performance is affected by the readout power, we perform QNDness measurements while sweeping the readout pulse's parameters. The pulse sequence is the same as in Fig. \ref{fig:qnd_vs_params}(a), however the duration $T_r$ and number of photons $\bar{n}_r$ of the two readout pulses are swept simultaneously. The value of $P_\text{QND}$ is computed for each parameter value, using the first definition. The errors $1-P_\text{QND}$ are plotted as function of $T_r$ and $\bar{n}_r$ in Fig. \ref{fig:qnd_vs_params}(b).

The results show a large plateau up to 300 photons with less than 4\% QNDness errors. Our optimal readout parameters fall within this region. The plateau is delimited by four limiting cases. At short pulse duration and at low pulse power, the SNR plummets and so the overlap between the $\ket{0}$ and $\ket{1}$ gaussians induces assignment errors for $m_1$ and $m_2$. At long readout duration, the finite $T_1$ starts inducing relaxation errors between $m_1$ and $m_2$, increasing $\proba(1,0)$. At very high readout powers, around 400 photons, the large readout photon population causes non-QND errors in the device. 

A slice of the colorplot is taken by fixing the readout duration to the optimal $T_r$, in Fig. \ref{fig:qnd_vs_params}(b). From this slice, the finer features of the central plateau can be seen. The QNDness errors actually present a small variation within this region and, as $\bar{n}_r$ increases, the errors also rise by small steps. These small variations seem to be the limiting factor for our QNDness at intermediate readout power.

\section{Discussion}
\label{sec:discussion}

For the post-selected fidelity without leakage, the current sample shows a total of $7.9\times10^{-3}$ readout errors, which can be divided into several phenomena. Due to an overlap of the $\ket{0}$ and $\ket{1}$ gaussians, we have $5\times10^{-4}$ SNR errors. The remaining $7.4\times10^{-3}$ are transition errors, from which only $2\times10^{-3}$ are expected to be relaxation errors due to the finite qubit lifetime $T_1=\SI{124.5}{\micro\second}$. The rest of the errors are due to qubit transitions induced by the readout power: $3.5\times10^{-3}$ of transitions from $\ket{1}$ to $\ket{0}$ and $1.9\times10^{-3}$ from $\ket{0}$ to $\ket{1}$. These induced transitions are not expected from the cos$\varphi$-coupling, which conserves the parity of the qubit's state (see Appendix \ref{apx:theory}.1). They could be due to high-power phenomena which are not mediated by the cos$\varphi$-coupling such as inelastic scattering \cite{Connolly2025}.



The current readout performance can be further improved by working on the different contributions to the readout's SNR. When the readout mode is driven on resonance and assuming the resonator has reached a steady state, the SNR can be expressed as \cite{Gambetta2008,Blais2021}:
\begin{align}
\label{eq:snr}
    \text{SNR}^2 = 4\eta T_r \bar{n}_r \frac{\kappa_r\chi_{qr}^2}{\kappa_r^2/4 + \chi_{qr}^2},
\end{align}
where $T_r$ (resp. $\bar{n}_r$) is the readout pulse's duration (resp. power, in number of readout photons), $\eta$ is the setup's quantum efficiency and $\kappa_r$ is the readout mode's losses. For our setup, the quantum efficiency is currently about 20\%, even though we are using a TWPA (see Appendix \ref{apx:setup}). By optimizing the setup, a quantum efficiency of 40\% should be achievable with a TWPA \cite{Spring2024}. Furthermore, the sample's parameters can be tuned to yield higher SNR. Note that the readout power $\bar{n}_r$ is limited by critical numbers of photons, which also depend on the circuit's parameters (see Appendix \ref{apx:theory}.3) so a simultaneous optimization of the entire expression $\bar{n}_r^\text{crit} \kappa_r\chi_{qr}^2/(\kappa_r^2/4 + \chi_{qr}^2)$ will be necessary. If the parameters and setup are optimized for high SNR, the benefits of high-power readout can be exploited further. The readout duration $T_r$ can be significantly shortened, making fast high-fidelity readout possible.

Furthermore, even though our readout is quite stable in the high-power regime, it is not perfectly insensitive to the effect of power. From the optimal point at 89 photons until around 300 photons, the QNDness errors increase weakly as a function of power, as shown in Section \ref{sec:readout}. This cannot be attributed to a resonant effect like MIST \cite{Sank2016}. Instead it could be attributed to a heating effect when the readout power increases due to a weak thermalization between the sample and the copper cavity. This is supported by the relatively high effective temperature of the qubit mode, of $\SI{72}{\milli\kelvin}$, around \SI{40}{\milli\kelvin} higher than the base temperature. A better thermalization of the sample will be beneficial to improve readout performance in the high-power regime.

\section{Current limitations of the high-power regime}
\label{sec:ncrit}





Interestingly, the standard critical number of photons introduced in the transverse-coupling readout scheme \cite{Blais2004,Gambetta_2006} is not a limit anymore for the $\cos\varphi$-coupling, since its cross-Kerr term is not perturbative. Indeed, in the standard readout scheme, that limitation stems from the large-detuning approximation. This specific critical number of photons is undefined for the $\cos\varphi$-coupling, but a similar study can be performed for the theoretical limits of the transmon molecule readout scheme's high-power regime.

The high-power limits are associated to the approximations made when derivating the cross-Kerr Hamiltonian Eq. (\ref{eq:ham_twomode}). Three approximations were made: the RWA, the linearization of the readout mode and the low-phase drop approximation. Each of these theoretical limits is detailed in Appendix \ref{apx:theory}.3 and the main conclusions are presented here. The RWA is performed both in the $\cos\varphi$-coupling and in the transverse coupling readout schemes. Its limit corresponds to a strong hybridization of the qubit's states due to the first non-RWA coupling terms. In our case, the critical number of photons derived from this is $\bar{n}_r^\text{RWA} = 2068$ photons. As for the linearization of the readout mode, it is also an approximation made in the standard readout scheme. The nonlinear effects are considered limiting when the threshold for the readout mode's bistability regime is reached. For the current anharmonicity $\alpha_r/(2\pi) = \SI{6.82}{\kilo\hertz}$, the bifurcation threshold is $\bar{n}_r^\text{bifurc} = \kappa_r/|3\sqrt{3}\alpha_r| = 505$ photons \cite{Ong2011,Dassonneville2023}. The last high-power limit is associated to the low phase-drop approximation \cite{Eichler2014,Eichler2014a}, used to Taylor expand the cosines of the $\cos\varphi$-coupling. It is specific to the transmon molecule readout scheme and results in the critical number of photons $\bar{n}_r^{\text{low} \varphi} = \sqrt{ E_J(1+2L_J/L_a) / ( 4 E_{C_a}) } / \sin^2(\theta) = 377$ photons. In the current sample, the low phase-drop approximation is the most limiting so $\bar{n}_r^\text{crit} = \bar{n}_r^{\text{low}\varphi} = 377$ photons. Note that this coincides with an abrupt increase of QNDness errors, as shown in Fig. \ref{fig:qnd_vs_params}(b). This number can be improved by reducing the readout mode's hybridization angle $\theta$ with flux-tuning.



The current transmon molecule sample can be compared to an equivalent standard readout scheme. The equivalent parameters are chosen so that both readout schemes display identical frequencies, qubit anharmonicity and dispersive shift. The chosen parameters are explained in Appendix \ref{apx:theory}.4. The resulting equivalent transverse readout scheme presents a critical number of photons $\bar{n}_\text{std}^\text{crit} = 26$, which prevents a large signal-to-noise ratio. In comparison, the critical number for the transmon molecule scheme is one order of magnitude higher. Furthermore, the standard scheme presents a short qubit relaxation time $T_1 \approx \SI{800}{\nano\second}$ due to Purcell effect. In contrast, the measured transmon molecule displays $T_1 = \SI{124}{\micro\second}$ without any Purcell filtering.

\section{Conclusion}
\label{sec:conclusion}

In summary, this work studies the high-power readout performance of the $\cos\varphi$-coupling readout scheme, implemented through a transmon molecule. The sample displays a readout fidelity of 99.21\% with post-selection, measured using 89 photons in the readout mode. This is a very high power compared to most high-fidelity measurements with a standard readout scheme. The QNDness of the readout is measured around 96.7\% at such powers.

A study of the sample's readout performance as a function of the power shows that there is a large region of intermediate to high readout powers where the QNDness errors stay low: up to 300 readout photons, the errors stay below 4\%. To put this in perspective, 300 readout photons are equivalent to more than 1200 times the qubit's energy and our readout mode is coupled to the qubit through the cos$\varphi$-coupling, which comes from a direct galvanic connection. Even so, there are less than 4\% errors induced by exchange of energy between the qubit and readout photons. 

This stability of the readout performance in the high-power regime is qualitatively explained by the study of the cos$\varphi$-coupling's theoretical high-power limitations. We can thus compute a novel critical number of photons $\bar{n}_r^\text{crit} = 377$ photons, associated to the validity of the coupling's Taylor expansion. These observations showcase the promising QNDness of the nonlinear cos$\varphi$-coupling for high-fidelity qubit readout.

\begin{acknowledgments}
The authors thank R. Dassonneville, E. Dumur, A. Petrescu, P. Leek and R. Vijay for insightful discussions.
This work benefited from a French government grant managed by the ANR agency under the ‘France 2030 plan’, with reference ANR-22-PETQ-0003 and BPI France AD ASTRA. O.B., C.M., V.M and F.D’E. acknowledge support from ANR OCTAVES (ANR-21-CE47-0007). T.R. acknowledges funding from the Generaci\'on de Conocimientos project PID2023-146531NA-I00 and the Ram\'on y Cajal program RYC2021-032473-I, financed by MCIN/AEI/10.13039/501100011033 and the European Union NextGenerationEU/PRTR.

C.M. and V.M. contributed equally to this work. V.M. optimized the sample's design and fabrication process for high-fidelity readout; C.M. further improved V.M.'s process to fabricate the final sample and conducted the experiments.
\end{acknowledgments}

\appendix

\section{Theory}
\label{apx:theory}

\subsection{Transmon molecule Hamiltonian}

The Hamiltonian for the transmon molecule circuit [green circuit in Fig. \ref{fig:circuit}(c)] is derived in \cite{Diniz2013,Dassonneville2020}. It is flux-dependent due to the superconducting loop formed by the two Josephson junctions and the inductance and takes the following form:
\begin{align}
    \begin{split}
    \hat{H}_\text{tm}&(\Phi_\text{ext})=4 \ECq \bnq^2-E_{Jq} \cos \left(\bphq\right) +4 \ECa \bna^2 \\
    -&2 E_J\left[\cos \left(\bpha\right)-\frac{L_J}{L_a(\Phi_\text{ext})} \left(\bpha - \frac{\Phi_\text{ext}}{2\varphi_0}\right)^2\right] \\
    -&2 E_J\left[\cos \left(\bphq\right)-1\right]\left[\cos \left(\bpha\right)-1\right],
    \end{split}
    \label{aeq:Htm_flux}
\end{align}
where $\Phi_\text{ext}$ is the magnetic flux passing through the superconducting loop, $\varphi_0=\hbar/(2e)$ is the reduced flux quantum, $\ECq=e^2/(4C_s)$ [resp. $\ECa=e^2/(8C_t+4C_s)$] is the qubit (resp. ancilla) mode's charging energy, $E_J$ is the individual Josephson junctions' energy, $E_{Jq}=2\EJ$ is the qubit mode's effective Josephson energy, $L_J=(\varphi_0)^2/E_J$ is the Josephson inductance, $\bnq$ (resp. $\bna$) is the qubit (resp. ancilla) mode's charge number operator and $\bphq$ (resp. $\bpha$) is the qubit (resp. ancilla) mode's phase operator. The inductance $L_a(\Phi_\text{ext})$ is flux-dependent since it is made from a chain of SQUIDs: $\La=\La^0/|\cos[\pi\Phi_\text{ext}/(\varphi_0A_\text{ratio})]|$, with $\La^0$ the inductance value at zero-flux and $A_\text{ratio}=28$ the ratio of areas between the transmon molecule's superconducting loop and the loop of the small SQUIDs composing the inductance. The Hamiltonian can be simplified at integer fluxes as follows:
\begin{align}
    \begin{split}\label{aeq:Htm}
    \hat{H}_\text{tm}=&4 \ECq \bnq^2-E_{Jq} \cos \left(\bphq\right) +4 \ECa \bna^2 \\
    &-2 E_J\left[\cos \left(\bpha\right)-\frac{L_J}{L_a(n)} \bpha^2\right] \\
    &-2 E_J\left[\cos \left(\bphq\right)-1\right]\left[\cos \left(\bpha\right)-1\right],
    \end{split}
\end{align}
where $L_a(n)$ is a function of $n$ the number of flux quanta trapped in the transmon molecule's superconducting loop. Since all measurements were performed at zero flux, we can consider that the inductance keeps the constant value $L_a^0$ and it will be simply noted $L_a$ in the following.

The coupling term between the two modes $-2\EJ \left[\cos(\bphq)-1\right] \left[\cos(\bpha)-1\right]$ comes from galvanic coupling and is purely nonlinear without any linear term. It is named $\cos\varphi$-coupling. By applying the low phase-drop approximation $\langle \hat{\varphi}_{q,a} \rangle \ll1$ \cite{Eichler2014,Eichler2014a}, the Hamiltonian is developed up to the fourth order, leading to an effective cross-Kerr coupling. We also introduce the annihilation operator $\hat{q}$ (resp. $\hat{a}$) for the qubit (resp. ancilla), as shown in \cite{Dassonneville2020}. The resulting equation, after RWA, can be written as follows:
\begin{align}
\begin{split}\label{aeq:Htm_taylor}
\hat{H}_\text{tm} /\hbar = &\omega_q\hat{q}^\dagger\hat{q} + \frac{\alpha_q} {2}\hat{q}^\dagger\hat{q}^\dagger \hat{q}\hat{q} + \omega_a\hat{a}^\dagger\hat{a} \\
&+ \frac{\alpha_a} {2}\hat{a}^\dagger\hat{a}^\dagger \hat{a}\hat{a}+ 2\chi_{qa} \hat{q}^\dagger\hat{q} \hat{a}^\dagger\hat{a},
\end{split}
\end{align}
where $\omega_q = \sqrt{4E_{Jq}\ECq}/\hbar + \alpha_q + \chi_{qa}$ (resp. $\omega_a = 4\sqrt{\EJ\ECa(1+2L_J/L_a)}/\hbar + \alpha_a + \chi_{qa}$) is the qubit's (resp. the ancilla's) frequency, $\alpha_q$ (resp. $\alpha_a$) is the qubit's (resp. the ancilla's) anharmonicity, and $\chi_{qa}$ is the strength of the cross-Kerr coupling between the qubit and ancilla. This strength can be expressed as $\chi_{qa}=-\sqrt{\ECq\ECa/(1+2\LJ/\La)}/\hbar$.

The resulting cross-Kerr term is ideal for QND readout of the qubit through the ancilla mode and its strength is independent of the detuning between the modes and can be adjusted by tuning the circuit parameters. No large-detuning approximation was applied to obtain this cross-Kerr term so the standard critical number of photons introduced in the transverse-coupling readout scheme \cite{Blais2004,Gambetta_2006} is not a limit anymore for the $\cos\varphi$-coupling. Furthermore the full development of the $\cos\varphi$-coupling only presents terms of the form $(\hat{q}^\dagger)^{2n}\hat{a}^{2m}$ and $\hat{q}^{2n}(\hat{a}^\dagger)^{2m}$, with $n$ and $m$ strictly positive integers. Consequently, the only transitions which can be induced by the $\cos\varphi$-coupling must conserve the parity for the number of transmon excitations and for the number of readout photons. Note that this reduces the number of allowed measurement-induced state transitions (MIST) when using the ancilla to measure the qubit. This property will be discussed in a future work.


\subsection{Hybridization of the readout mode}


The previous subsection showed that the transmon molecule implements a transmon which can be measured through another circuit mode, called ancilla. The ancilla mode has promising characteristics for a readout mode but it cannot be used as such because it is uncoupled to the exterior in our present description. In order to engineer a readout scheme using the transmon molecule, we add a third mode to the system: a resonator with strong transverse coupling to the ancilla, and decoupled from the qubit. From this specific configuration, the qubit mode is intrinsically protected from Purcell effect. The equivalent electric circuit for the readout scheme, using a transmon molecule and a transmission microwave resonator, is shown in Fig. \ref{fig:circuit}(c). The full Hamiltonian now becomes:
\begin{align}
\label{aeq:ham_full}
    \hat{H}_{\text{tot}} &= \hat{H}_{\text{tm}}
     + \hbar\wc\hat{c}^\dag\hat{c} - \hbar\ga\left(\hat{a}-\hat{a}^\dag\right) \left(\hat{c}-\hat{c}^\dag\right),
\end{align}
where the resonator is defined by its annihilation operator $\hat{c}$ and resonant frequency $\wc$ and the strength of the ancilla-resonator transverse coupling is noted $\ga$. After applying the low phase-drop approximation and the RWA, the full Hamiltonian can be written as follows:
\begin{align}
    \begin{split}\label{aeq:Htot}
    \hat{H}_\text{tot}&/\hbar = \omega_q\hat{q}^\dagger\hat{q} + \frac{\alpha_q} {2}\hat{q}^\dagger\hat{q}^\dagger \hat{q}\hat{q} \\
    &+ \omega_a\hat{a}^\dagger\hat{a} + \frac{\alpha_a} {2}\hat{a}^\dagger\hat{a}^\dagger \hat{a}\hat{a}+ 2\chi_{qa} \hat{q}^\dagger\hat{q} \hat{a}^\dagger\hat{a} \\
    &+ \wc\hat{c}^\dag\hat{c} + \ga\left(\hat{a} \hat{c}^\dag +\hat{a}^\dag \hat{c} \right) .
    \end{split}
\end{align}
Due to the strong coupling between them, the ancilla mode and resonator hybridize. By approximating the ancilla mode as linear, we change basis as shown in \cite{Dassonneville2020} and get two hybrid modes. These modes are hybrids between solid-state excitations (of the ancilla mode) and photons (of the resonator mode) so they are polariton modes. As such, the polaritons inherit the cross-Kerr coupling to the qubit (from the ancilla) as well as a coupling to the exterior (from the resonator). This means they can both be used as readout modes. They are noted $u$ and $l$ for ``upper polariton'' and ``lower polariton'', respectively. The resulting three-mode Hamiltonian after RWA is as follows:
\begin{align}
    \begin{split}\label{aeq:Hpol}
    \hat{H}_\text{pol}&/\hbar = \omega_q\hat{q}^\dagger\hat{q} + \frac{\alpha_q} {2}\hat{q}^\dagger\hat{q}^\dagger \hat{q}\hat{q} \\
    &+ \omega_l\hat{l}^\dag\hat{l} + \frac{\alpha_l} {2}\hat{l}^\dagger\hat{l}^\dagger \hat{l}\hat{l} + 2\chi_{ql} \hat{q}^\dagger\hat{q} \hat{l}^\dagger\hat{l} \\
    &+ \omega_u\hat{u}^\dag\hat{u} + \frac{\alpha_u} {2}\hat{u}^\dagger\hat{u}^\dagger \hat{u}\hat{u} + 2\chi_{qu} \hat{q}^\dagger\hat{q} \hat{u}^\dagger\hat{u} \\
    &+ 2\chi_{ul} \hat{l}^\dagger\hat{l} \hat{u}^\dagger\hat{u} ,
    \end{split}
\end{align}
where the operators $\hat{l}=\cos(\theta)\hat{a}+\sin(\theta)\hat{c}$ and $\hat{u}=-\sin(\theta)\hat{a}+\cos(\theta)\hat{c}$ are the annihilation operators for the lower and upper polariton, respectively, with $\theta=(1/2)\arctan[2g_{ac}/(\omega_c-\omega_a)]$ the mixing angle. The lower (resp. upper) polariton's resonant frequency is $\omega_l=\sin^2(\theta)\omega_c+\cos^2(\theta)\omega_a+\sin(2\theta)g_{ac}$ (resp. $\omega_u=\cos^2(\theta)\omega_c+\sin^2(\theta)\omega_a-\sin(2\theta)g_{ac}$) and its anharmonicity is $\alpha_l= \cos^4(\theta)\alpha_a$ (resp. $\alpha_u= \sin^4(\theta)\alpha_a$). We also introduced the strength $\chi_{ql}= \cos^2(\theta) \chi_{qa}$ (resp. $\chi_{qu}= \sin^2(\theta) \chi_{qa}$) of the cross-Kerr coupling between the qubit and lower (resp. upper) polariton. There is also a cross-Kerr coupling between the polaritons, of strength $\chi_{ul} = 2\cos^2(\theta) \sin^2(\theta) \alpha_a$. Note that, for this work's sample, the bare cavity's frequency is higher than the bare ancilla's, so the upper polariton is more cavity-like and the lower polariton is more ancilla-like at zero-flux. The losses for the two polaritons can also be written as a function of the bare mode's losses: $\kappa_l = \kappa_c \sin^2(\theta) + \kappa_a \cos^2(\theta)$ and $\kappa_u = \kappa_c \cos^2(\theta) + \kappa_a \sin^2(\theta)$ for the lower and upper polariton, respectively.

The Hamiltonian (\ref{aeq:Hpol}) can be simplified further by considering that only one of the polaritons is used for qubit readout. We choose the upper polariton, which is cavity-like and thus more linear. If we consider that driving this polariton does not induce significant population of the lower polariton, then we can consider that the unused polariton remains in the vacuum state and we neglect its energy terms as well as the inter-polariton cross-Kerr. The simplified two-mode Hamiltonian can be written as follows:
\begin{align}
    \begin{split}\label{aeq:Hpol_disp}
    \frac{\hat{H}_{\text{reduced}}}{\hbar} &= \wq \hat{q}^\dagger \hat{q} + \frac{\alpha_q} {2} \hat{q}^\dagger\hat{q}^\dagger \hat{q}\hat{q} \\
    &+\omega_r\hat{c}_r^\dag\hat{c}_r +  \frac{\alpha_r} {2} \hat{c}_r^\dagger\hat{c}_r^\dagger \hat{c}_r\hat{c}_r \\
    &+ 2\chi_{qr} \hat{q}^\dagger \hat{q} \hat{c}_r^\dag\hat{c}_r,
    \end{split}
\end{align}
where $\omega_r=\omega_u$ and $\chi_{qr}=\chi_{qu}$. This Hamiltonian corresponds to Eq. (\ref{eq:ham_twomode}) in the main text. It is a dispersive Hamiltonian which can be used for qubit readout and it does not rely on any high-detuning approximation, different from the case of transverse coupling. Furthermore, the cross-Kerr term is not perturbative, since it comes from the $\cos\varphi$-coupling which is itself purely nonlinear.

\subsection{Critical number of photons for the transmon molecule}

Since the transmon molecule readout scheme does not rely on any high-detuning approximation, the standard critical number of photons becomes meaningless. There are thus other approximations which become limiting in the high-power regime. The three main assumptions made for attaining the ideal cross-Kerr coupling term in Eq. (\ref{aeq:Hpol_disp}) are the RWA, the linearization of the readout mode and the low phase-drop approximation. Each of them leads to a different critical number of photons.

\subsubsection{High-power limit of the RWA}

Without the RWA, the readout Hamiltonian from Eq. (\ref{aeq:Hpol_disp}) takes the form:
\begin{align}
    \begin{split}\label{aeq:Hpol_disp_noRWA}
    \hat{H}_\text{pol} /\hbar&=\omega_q\hat{q}^\dagger\hat{q} + \frac{\alpha_q}{12} \left(\hat{q}+\hat{q}^\dag\right)^4 \\
    &+ \omega_r\hat{c}_r^\dagger\hat{c}_r+\frac{\chi_{qr}}{2} \left( \hat{q} + \hat{q}^\dagger\right)^2 \left(\hat{c}_r + \hat{c}_r^\dagger\right)^2 .
    \end{split}
\end{align}
Under the RWA, the only coupling term that is kept from $\left( \hat{q} + \hat{q}^\dagger\right)^2 \left(\hat{c}_r + \hat{c}_r^\dagger\right)^2$ is the cross-Kerr one. However, when the number of photons is increased, some of the non-RWA coupling terms become significant and must be taken into account. With the current circuit parameters, the first non-RWA terms are $(\hat{q}^\dag)^2\hat{c}_r^\dag\hat{c}_r$ and $\hat{q}^2\hat{c}_r^\dag\hat{c}_r$, which rotate at $2\wq/(2\pi)=\SI{4}{\giga\hertz}$. 
When accounting for these terms, the dressed eigenstates can be computed perturbatively for small $\chi_{qr}$:
\begin{align}
\label{aeq:dressed_eigstates}
\begin{cases}
    \overline{\ket{0,n}} &= \frac{1}{N_{0,n}} \left[ \ket{0,n} + \frac{\chi_{qr}\sqrt{2}n}{\bar{E}_{0,n} - \bar{E}_{2,n}} \ket{2,n} \right] \\
    \overline{\ket{1,n}} &= \frac{1}{N_{1,n}} \left[ \ket{1,n} + \frac{\chi_{qr}\sqrt{6}n}{\bar{E}_{1,n} - \bar{E}_{3,n}} \ket{3,n} \right] 
\end{cases}\,,
\end{align}
with the notation $\ket{k,n}$ (resp. $\overline{\ket{k,n}}$) for a bare (resp. dressed) eigenstate with $k$ transmon excitations and $n$ readout photons. The eigenenergy associated to a dressed eigenstate $\overline{\ket{k,n}}$ is noted $\bar{E}_{k,n}$. The eigenvectors are normalized by the constants $N_{0,n}^2 = 1 + \left(\frac{\chi_{qr}\sqrt{2}n}{\bar{E}_{0,n} - \bar{E}_{2,n}}\right)^2$ 
and $N_{1,n}^2 = 1 + \left(\frac{\chi_{qr}\sqrt{6}n}{\bar{E}_{1,n} - \bar{E}_{3,n}}\right)^2$. The perturbative term in each dressed eigenstate scales with the number of readout photons $n$ and becomes significant in the high power regime. A critical number of photons can be defined for when the perturbative term reaches the same probability amplitude as the non-perturbative term. This is more limiting in the case of eigenstate $\overline{\ket{1,n}}$ since the perturbative term is larger than for $\overline{\ket{0,n}}$. The resulting critical number of photons is:
\begin{align}
\label{aeq:ncrit_RWA}
    \bar{n}_r^\text{RWA} = \left| \frac{\omega_{13}}{\sqrt{6}\chi_{qr}} \right|=2068,
\end{align}
where $\omega_{13}/(2\pi) = \SI{3.90}{\giga\hertz}$ is the frequency of the transmon's $\ket{1}$ to $\ket{3}$ transition. The resulting critical number of photons is not limiting with the current circuit's parameters.

The critical number $\bar{n}_r^\text{RWA}$ indicates the threshold around which the non-RWA terms $(\hat{q}^\dag)^2\hat{c}_r^\dag\hat{c}_r$ and $\hat{q}^2\hat{c}_r^\dag\hat{c}_r$ stop being negligible. With our current parameters, these are the slowest-rotating non-RWA terms. However, in situations where $\omega_q > |\omega_r-\omega_q|$, the dominant non-RWA terms become $(\hat{q}^\dag)^2\hat{c}_r^2$ and $\hat{q}^2(\hat{c}_r^\dag)^2$. A different critical number can be associated to these terms. Following an analogous derivation to that of $\bar{n}_r^\text{RWA}$, we find:
\begin{align}
\label{aeq:ncrit_RWA2}
    \bar{n}_r^\text{RWA,2} = \frac{2}{\sqrt{6}} \left| \frac{2\omega_r - \omega_{13} }{\chi_{qr}} \right|=3595.
\end{align}
This critical number of photons is far from limiting with our sample's parameters. Note that $\bar{n}_r^\text{RWA,2}$ depends on the detuning between qubit and readout mode $\Delta = |\omega_r - \omega_q|$. If the qubit mode is approximated as an anharmonic oscillator, then $\omega_{13} \approx 2\omega_q + 3\alpha_q$ and so Eq. (\ref{aeq:ncrit_RWA2}) becomes $\bar{n}_r^\text{RWA,2} = (2/ \sqrt{6}) | (2\Delta - 3\alpha_q) / \chi_{qr} |$.


\subsubsection{Nonlinearity of the readout mode}

The linearization of the ancilla mode is performed when writing the transmon molecule readout scheme Hamiltonian in the hybridized polariton basis. In reality, the bare ancilla has an anharmonicity $\alpha_a/(2\pi) = \SI{-1.29}{\mega\hertz}$, which results in a anharmonicity $\alpha_r/(2\pi) = \SI{-6.82}{\kilo\hertz}$ for the polariton used as readout mode. This results in nonlinear behavior in the high-power regime. At very high-powers, the readout mode can bifurcate and reach a bistability regime \cite{Ong2011,Dassonneville2023}. This was previously used for performing a latching readout \cite{Dassonneville2023}. The threshold for the bistability regime is given by:
\begin{align}
\label{aeq:ncrit_bifurc}
    \bar{n}_r^\text{bifurc} = \frac{\kappa_r}{\left| 3\sqrt{3}\alpha_r \right|} = 505.
\end{align}
Note that this phenomenon is not intrinsically non-QND, since the bistability region can be avoided by tuning the readout frequency \cite{Dassonneville2023}. However, it should be accounted for when analyzing the experimental results since the bistable behavior of the readout mode can lead to misinterpreting the results.

\subsubsection{Low phase-drop approximation}

The low phase-drop approximation $\langle \hat{\varphi}_{q,a} \rangle \ll1$ \cite{Eichler2014,Eichler2014a} is applied to develop the $\cos\varphi$-coupling up to the fourth order, leading to the cross-Kerr coupling term. However, it does not hold when the phase fluctuations become too large. This becomes an issue for the bare ancilla when its population is increased too much. An upper bound can be set for the ancilla population by analogy with the critical number of photons criterion used in \cite{Eichler2014,Eichler2014a,Frattini2018,Planat2019}. When the ancilla is in a coherent state of average population $\bar{n}_a$, then the condition $\langle \hat{\varphi}_a^2 \rangle \ll 1$ can be rewritten as:
\begin{align}
\label{aeq:ncrit}
\bar{n}_a \ll \bar{n}_a^{\text{low}\varphi} = \frac{1}{2} \sqrt{ \frac{E_J (1+ \frac{2L_J}{L_a})} {E_{C_a}} }.
\end{align}
This gives an estimation of how many bare ancilla mode excitations are allowed before the fourth-order development of the $\cos \left(\bpha\right)$ stops being valid. Above this limit, the ancilla mode presents nonlinear behavior leading to a self-Kerr frequency shift. In this regime, the exact complete expression of the cos$\varphi$-coupling must be considered, which leads to non-QND effects and resonant transitions.

From the critical number of bare ancilla excitations, a critical number of readout photons can be derived:
\begin{align}
\begin{split}
\label{aeq:ncrit_pol}
\bar{n}_r^{\text{low}\varphi} &= \frac{\bar{n}_a^{\text{low}\varphi}} {\sin^2 (\theta)} \\
&= \frac{1}{2 \sin^2(\theta)} \sqrt{ \frac{E_J (1+ \frac{2L_J}{L_a})} {E_{C_a}} } = 377 .
\end{split}
\end{align}
This represents the limit when the cos$\varphi$-coupling cannot be approximated as cross-Kerr anymore, due to the high number of excitations in the hybrid readout mode. With the current circuit parameters, the low phase-drop approximation is the most limiting assumption in the high-power regime.  Thus, for this sample, the critical number of photons is defined as $\bar{n}_r^\text{crit} = \bar{n}_r^{\text{low}\varphi} = 377$ photons. Note that this number of photons can be adjusted and increased by tuning the circuit parameters and the applied magnetic flux.

\subsection{Equivalent transverse-coupling readout scheme}

To highlight the benefits of the cos$\varphi$-coupling, the sample can be compared to an equivalent transversely-coupled-transmon. The standard readout scheme uses the Hamiltonian 
\begin{align}
\begin{split}
\label{aeq:std_eq}
\hat{H}^\text{eq}_\text{std}/\hbar&= \omega_q^\text{eq} \hat{q}^\dag \hat{q} + \frac{\alpha_q^\text{eq}}{2} \hat{q}^\dag \hat{q}^\dag \hat{q} \hat{q} \\
&+\omega_c^\text{eq}\hat{c}^\dagger\hat{c}+ g_x^\text{eq} \left( \hat{q} \hat{c}^\dag + \hat{q}^\dag \hat{c} \right) ,
\end{split}
\end{align}
where $\omega_q^\text{eq}$ is the transmon's resonant frequency, $\hat{q}$ is its annihilation operator, $\omega_c^\text{eq}$ is the resonator's frequency, $\hat{c}$ is its annihilation operator and $g_x^\text{eq}$ is the strength of the transverse coupling between the transmon and the resonator. To get the same bare spectra for both Hamiltonians, we choose $\omega^\text{eq}_q=\omega_q$ and $\omega^\text{eq}_c=\omega_r$. The coupling constant $g_x^\text{eq}$ is chosen to equate the dispersive shifts $\chi^\text{eq}=\chi_{qr}$, using the formula from \cite{Koch2007}: $g_x^\text{eq}=\sqrt{\chi_{qr}(\omega_q^\text{eq}-\omega_c^\text{eq})(\omega_q^\text{eq}+\alpha_q^\text{eq}-\omega_c^\text{eq})/ \alpha_q^\text{eq}}$, where $\alpha_q^\text{eq}= \alpha_q$ is the equivalent transmon's anharmonicity. For this equivalent readout scheme, the required transverse coupling strength is $g_x^\text{eq}/(2\pi)=\SI{515}{\mega\hertz}$. This leads to a very short qubit relaxation time due to Purcell effect: $T_1= (\omega_q^\text{eq}-\omega_c^\text{eq})^2/[\kappa_\text{out}(g_x^\text{eq})^2] \approx \SI{800}{\nano\second}$. Due to this intrinsic limitation of the standard transmon readout scheme, a Purcell filter stage is necessary \cite{Swiadek2024,Spring2024}. A second limitation of this coupling readout is the low critical number of photons $\bar{n}_\text{std}^\text{crit}=(\omega_q^\text{eq}-\omega_c^\text{eq})^2/(2g_x^\text{eq})^2=26$, which prevents a large signal-to-noise ratio. In comparison to $\bar{n}_\text{std}^\text{crit}$, the critical number of photons for the cos$\varphi$-coupling is one order of magnitude higher.


\section{Experimental setup}
\label{apx:setup}

\subsection{Full microwave setup}

\begin{figure}
\centering
\includegraphics[width=1\columnwidth]{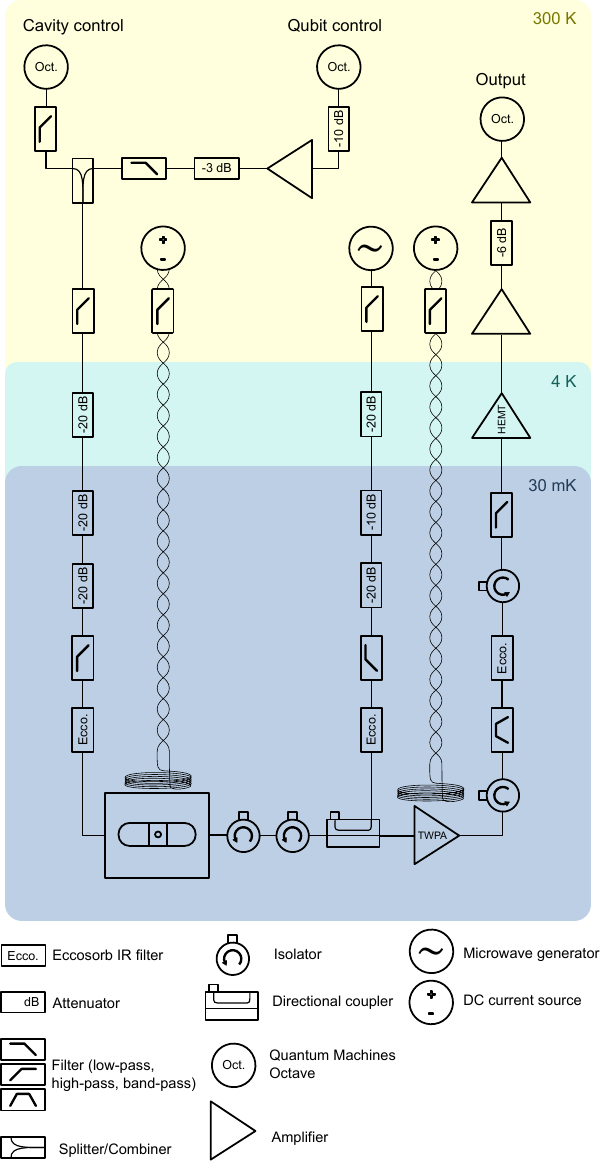}
\caption{\label{figapx:mwsetup} Full microwave setup. }
\end{figure}

The full microwave setup is shown in Fig. \ref{figapx:mwsetup}. The cavity control and qubit control pulses are generated by a Quantum Machines OPX+ and up-converted by a Quantum Machines Octave. The two control signals are then combined and sent to the cavity through a common microwave line. The cavity's output signal is amplified at base temperature by a homemade TWPA with ``reverse-Kerr'' phase-matching \cite{Planat2020,Ranadive2022}, which is pumped through a directional coupler. The TWPA is flux-biased by a superconducting coil connected to a Keysight B2902A current source. The pump tone is generated by an HP83630A microwave source. After the TWPA, the signal is amplified by a HEMT at 4K and then amplified at room temperature. The signal then gets down-converted by the Quantum Machines Octave and is acquired by the Quantum Machines OPX+.

Note that the low-temperature microwave setup presents several isolators and filters on the output line, in order to ascertain that we have no effect on the qubit from the TWPA's pump and from the TWPA's idler tone. However this has the effect of limiting the quantum efficiency, measured around 20\%. This setup could be further optimized for high-fidelity readout measurements, for example by using broader-band isolators and less filters on the output line.

\subsection{Cavity characteristics}

The 3D resonator used is a rectangular cavity made of OFHC copper, with dimensions $5\times24.5\times\SI{35}{\milli\meter}$. We use its $\text{TE}_{101}$ mode for the transmon molecule readout scheme. The cavity can be probed in transmission through an input and an output port. The ports' couplings were adjusted and measured at room temperature as $\kappa_\text{in}/(2\pi)=\SI{0.153}{\mega\hertz}$ and $\kappa_\text{out}/(2\pi)=\SI{13.0}{\mega\hertz}$. Those couplings were tuned so that $\kappa_\text{out}\gg\kappa_\text{in}$. A superconducting coil is wound around the cavity to allow for flux biasing. The coil is biased by a Keysight B2902A current source. The cavity and coil are placed inside a mu-metal half-cylindrical shield with an IR-absorbing coating on the inside.

\section{Fabrication recipe}
\label{apx:fab}

The fabrication recipe used is close to the one from Ref. \cite{Dassonneville2020} although several steps were improved upon. The sample is made from a high-resistivity \ce{Si}$\left(100\right)$ 2 inch wafer. At the start of the process, just before spin-coating, the wafer is cleaned using a buffer-oxide etchant (BOE) 1:7 solution. This removes impurities from the sample to avoid two-level systems (TLS), and removes the silicon oxide to minimize dielectric losses.

The Josephson junctions are made using two different techniques: bridge-free technique \cite{Lecocq2011} for the chain of SQUIDs which we use as an inductor, Dolan bridge technique \cite{Dolan1977} for the smaller individual junctions in the circuit. For that, two layers of electronic resist are spincoated: first a \SI{750}{\nano\meter} layer of PMMA/MAA 8\% (ARP 617.14/600.07), then a layer of \SI{270}{\nano\meter} of PMMA 950K 4\% (ARP.679.04). Then the junctions and the circuit are patterned using \SI{80}{\kilo\electronvolt} electron-beam lithography (NanoBeam Ltd.).

The mask is developed using cold development \cite{Hu2004,Cord2007,Ocola2006}. With this technique, the smaller structures, such as Josephson junctions, show cleaner and more defined edges. This makes the junctions' surface and Josephson energy more predictable and more reproducible. The developer is a solution of IPA/$\ce{H}_2\ce{O}$ 3:1, cooled to $1\pm\SI{0.1}{\celsius}$. The wafer is dipped for \SI{60}{\second} in the cold developer and then rinsed in cold water for about \SI{30}{\second} and dried. After the development, we perform a \SI{20}{\second} oxygen RIE (Reactive Ion Etching) descum at \SI{10}{\watt}.

For the aluminum evaporation, the sample is placed in an electron-beam evaporation system (Plassys). The chamber and loadlock are pumped overnight and then a titanium flash is performed to achieve a vacuum of around $\SI{2.5e-8}{\milli\bar}$ in the chamber. Then a \SI{20}{\nano\meter} layer of aluminum is evaporated at \SI{0.1}{\nano\meter\per\second}, at an angle of \SI{-35}{\degree}. An oxidation is performed by filling the chamber with $\ce{O}_2$ at a pressure of \SI{5}{\milli\bar} for 5 minutes. This is followed by pumping and a titanium flash to reach pressures around $\SI{3.5e-8}{\milli\bar}$. The second aluminum evaporation is then performed, to deposit a \SI{50}{\nano\meter} layer at \SI{0.1}{\nano\meter\per\second}, at an angle of \SI{+35}{\degree}. Finally, a second oxidation is performed for capping, with the same parameters as the first. This capping step improves the stability and reproducibility of the Josephson junctions' energy.

Lift-off can then be performed by leaving the wafer overnight in a beaker of NMP at \SI{80}{\celsius} and then rinsing with acetone, ethanol, IPA and drying. The wafer is diced into $5\times\SI{6.8}{\milli\meter}$ chips using a Disco$^\text{TM}$ DAD 321 Automatic Dicer.

\begin{figure}
\centering
\includegraphics[width=1\columnwidth]{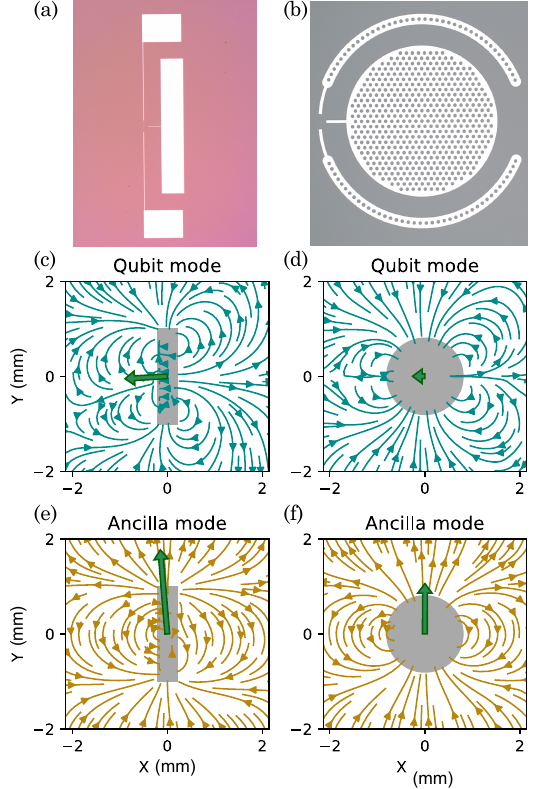}
\caption{\label{fig:hfss_simu} (a),(b) Optical microscopy pictures of the rectangular and circular designs, respectively. (c),(d),(e),(f) Simulated field distributions of the qubit and ancilla modes for an either rectangular or circular transmon molecule geometry. The grey shapes represent the transmon molecule's outline in each case and the green arrows indicate the total field magnitude, integrated over the entire chip. }
\end{figure}

\section{Sample optimization}
\label{apx:optim}

Three main parameters of the sample were optimized to get better readout performance than in Ref. \cite{Dassonneville2020}: the qubit-resonator detuning $\Delta_{qc}$ was increased, the qubit's dipole moment magnitude $|\bm{d}_q|$ was reduced and the ancilla mode's anharmonicity $\alpha_a$ was lowered. The dipole moment $\bm{d}_q$ was modified by changing the sample's design whereas $\Delta_{qc}$ and $\alpha_a$ were varied by tuning the circuit's electrical parameters.

\begin{figure}
\centering
\includegraphics[width=1\columnwidth]{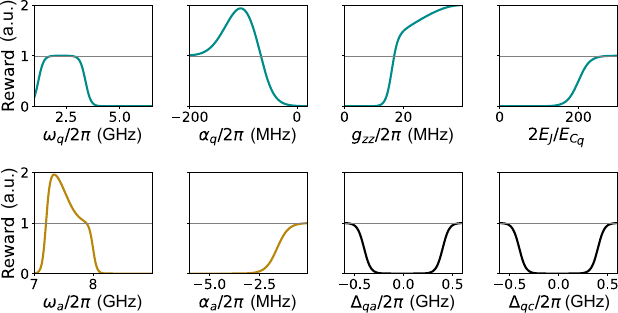}
\caption{\label{figapx:reward_fxs} Reward functions used for optimizing the circuit's parameters. }
\end{figure}

\subsection{Sample design}


As explained in the Section \ref{sec:implementation}, misaligning the transmon molecule and the 3D resonator can lead to remnants of transverse coupling and Purcell effect. In order to be less sensitive to misalignment, the design of the new generation of samples was changed to a circular shape, inspired by the concentric transmons \cite{Braumueller2016,Rahamim2017}. Since the transverse coupling strength is proportional to the magnitude of the dipole moment, the circular shape is meant to reduce the magnitude of the dipole moment of the qubit mode. At the same time, the dipole moment of the ancilla mode should keep a large magnitude so that it hybridizes with the resonator mode.

Eigenmode simulations were performed using Ansys HFSS, showing the field distribution of the qubit mode and the ancilla mode for the previous rectangular design \cite{Dassonneville2020} as well as the current circular design. All the modes are simulated for \SI{1}{\joule}oule of stored energy so that the results are comparable. Both simulated designs also have the same resonant frequencies: here, the qubit mode is around \SI{2}{\giga\hertz} and the ancilla mode is around \SI{3.5}{\giga\hertz}. From the simulations, we can compute the field distribution and the total average electric field generated by the eigenmode, as plotted in Fig. \ref{fig:hfss_simu}(c,d,e,f).

The generated fields shown in the results confirm the orthogonality of the qubit and ancilla modes' dipole moments, as mentioned in Section \ref{sec:implementation}. Furthermore, the magnitude of the qubit mode's generated field is visibly smaller in the case of a circular transmon molecule. To be more quantitative, we compare the ratio $|\bm{d}_q|/|\bm{d}_a|$, where $\bm{d}_q$ (resp. $\bm{d}_a$) is the qubit's (resp. the ancilla's) dipole moment. We find for the rectangular geometry $|\bm{d}_q|/|\bm{d}_a|=0.50$ and for the circular geometry $|\bm{d}_q|/|\bm{d}_a|=0.25$. This means that, by changing the geometry to a circular one, we can decrease $|\bm{d}_q|$ by 50\% while keeping a constant $|\bm{d}_a|$. This can be understood from the vector fields: in the circular geometry, the qubit mode generates mostly locally radial components, which compensate each other when averaged over the whole sample. This reduction of the qubit's dipole moment magnitude results in a 50\% reduction of the remnants of transverse coupling in case of a misalignment of the sample. In turn, this is equivalent to dividing the Purcell decay rate by a factor 4.

\begin{table*}
\centering
\begin{tabular}{c@{\hspace{2em}}c@{\hspace{2em}}c@{\hspace{2em}}c} 
\toprule
\toprule
Parameter name & Previous work's sample \cite{Dassonneville2020}  & Target values & Current work's sample \\
\toprule
$\wq/(2\pi)$ & \SI{6.284}{\giga\hertz} & \SI{2.78}{\giga\hertz} & \SI{2.0332}{\giga\hertz} \\
$\wa/(2\pi)$ & \SI{7.78}{\giga\hertz} & \SI{7.35}{\giga\hertz} & \SI{6.492}{\giga\hertz} \\
$\omega_c/(2\pi)$ & \SI{7.169}{\giga\hertz} &  & \SI{7.23}{\giga\hertz} \\
$\omega_u/(2\pi) = \omega_r/(2\pi)$ & \SI{7.911}{\giga\hertz} &  & \SI{7.290}{\giga\hertz} \\
$\omega_l/(2\pi)$ & \SI{7.038}{\giga\hertz} &  & \SI{6.432}{\giga\hertz} \\

$\alpha_q/(2\pi)$ & \SI{-88}{\mega\hertz} & \SI{-77}{\mega\hertz} & \SI{-73.1}{\mega\hertz} \\
$\alpha_a/(2\pi)$ & \SI{-13.5}{\mega\hertz} & \SI{-1.3}{\mega\hertz} & \SI{-1.29}{\mega\hertz} \\
$\alpha_u/(2\pi)=\alpha_r/(2\pi)$ & \SI{-11.6}{\mega\hertz} &  & \SI{-0.00682}{\mega\hertz} \\
$\alpha_l/(2\pi)$ & \SI{-0.0713}{\mega\hertz} & & \SI{-1.11}{\mega\hertz} \\
$\chi_{qa}/(2\pi)$ & \SI{-34.5}{\mega\hertz} & \SI{-20.2}{\mega\hertz} & \SI{-10.4}{\mega\hertz} \\
$\chi_{qu}/(2\pi) = \chi_{qr}/(2\pi)$ & \SI{-28.5}{\mega\hertz} &  & \SI{-0.77}{\mega\hertz} \\
$\chi_{ql}/(2\pi)$ & \SI{-4.5}{\mega\hertz} &  & \SI{-9.6}{\mega\hertz} \\
$g_{ac}/(2\pi)$ & \SI{295}{\mega\hertz} &  & \SI{224}{\mega\hertz} \\

$\kappa_u/(2\pi) = \kappa_r/(2\pi)$ & \SI{7.1}{\mega\hertz} &  & \SI{17.9}{\mega\hertz} \\
$\kappa_l/(2\pi)$ & \SI{11.8}{\mega\hertz} &  & \SI{2.84}{\mega\hertz} \\
$\theta$ & \SI{0.384}{rad} &  & \SI{0.273}{rad} \\

\midrule
$C_s$ & \SI{110}{\femto\farad} & \SI{125}{\femto\farad} & \SI{132}{\femto\farad} \\
$C_t$ & \SI{59.6}{\femto\farad} & \SI{87}{\femto\farad} & \SI{96.6}{\femto\farad} \\
$E_J/(2\pi\hbar)$ & \SI{29.2}{\giga\hertz} & \SI{7.3}{\giga\hertz} & \SI{3.84}{\giga\hertz} \\
$L_a$ & \SI{5.32}{\nano\henry} & \SI{2.66}{\nano\henry} & \SI{3.85}{\nano\henry} \\
$E_{Jq}/\ECq$ & 665 & 189 & 105 \\

\midrule
$T_1$ & \SI{3.3}{\micro\second} &  & \SI{124.5}{\micro\second} \\
$T_2*$ & \SI{3.2}{\micro\second} &  & \SI{10.6}{\micro\second} \\
$T_2^E$ &  &  & \SI{22.6}{\micro\second} \\

\bottomrule
\bottomrule
\end{tabular}
\caption{Sample parameters for the previous work and the current work. The target values indicate the set of optimized parameters that were aimed for during the fabrication of this work's sample.}
\label{tab:target_params}
\end{table*}

\subsection{Circuit parameters}



The circuit's electrical parameters must be tuned to increase $\Delta_{qc}$ and reduce $\alpha_a$. However, there are many constraints to be considered simultaneously, so we start from the previous set of parameters from \cite{Dassonneville2020}, shown in Table \ref{tab:target_params}, and vary them to improve the sample's performance.

There are two main reasons for changing the circuit parameters. First, in order to eliminate any remnants of Purcell effect, we choose to increase the qubit-resonator detuning $|\Delta_{qc}| = \left|\wq-\wc\right|$, by lowering the qubit's resonant frequency. Note that this will not affect the strength of the cross-Kerr coupling. A second goal is to improve the readout performance of the new samples, so we decrease the ancilla mode's anharmonicity $\alpha_a$, thus aiming for a very linear readout mode. This allows larger numbers of photons in the readout mode before reaching the nonlinear regime\cite{Dassonneville2023}.

The transmon molecule has 4 different circuit parameters that we can vary, shown in the circuit scheme Fig. \ref{fig:circuit}(a): $E_J$, $C_s$, $C_t$ and $L_a$. These parameters can tune the properties of the qubit mode and ancilla mode. The target parameters should yield a larger qubit-cavity detuning $|\Delta_{qc}|$ and a lower ancilla anharmonicity $\alpha_a$ but, aside from these goals, several other constraints must be verified while varying the parameters. The ancilla mode's frequency $\omega_a$ should stay close to $\omega_c/(2\pi)=\SI{7.3}{\giga\hertz}$. The qubit mode's anharmonicity should stay around \SI{-100}{\mega\hertz}, otherwise it cannot be approximated as a qubit. The ratio $E_{Jq}/E_{Cq}$ should stay above the 100 range to prevent charge noise. The qubit-ancilla coupling should be larger than \SI{10}{\mega\hertz} to have a big enough dispersive shift. Finally, it is preferable to avoid collisions between the qubit's 02 transition and the ancilla and cavity modes, so the detunings $\Delta_{02,a}=\left|\omega_{02}-\omega_{a}\right|$ and $\Delta_{02,c}=\left|\omega_{02}-\omega_{c}\right|$ are kept above \SI{300}{\mega\hertz}.

These constraints are translated into reward functions for optimization of the parameters, as shown in Fig. \ref{figapx:reward_fxs}. They are computed as sums or products of exponential, sigmoid and gaussian functions, in order to guarantee their smoothness. A standard gradient ascent algorithm using these reward functions yields the target parameters shown in Table \ref{tab:target_params}.

\section{Readout optimization}
\label{apx:ro_optim}

There are many pulse parameters to optimize in order to get the sample's maximal readout fidelity. The $\pi$-pulse's frequency and power are calibrated beforehand. Then the readout frequency $\omega_d$ is chosen with the TWPA off, to maximize the SNR between the pointer states for $\ket{0}$ and $\ket{1}$. The optimal drive frequency is $\omega_d=\omega_r$. Once $\omega_r$ is fixed, the TWPA pump's frequency and power are tuned to get maximal SNR improvement at this chosen frequency. The TWPA is kept on for all the following steps.

Then the duration and power of the pulses from the sequence of Fig. \ref{fig:fid}(a) are tuned. First, the parameters of the pre-pulse are varied to calibrate the state preparation. The duration $T_r^\text{pre}$ and power $\bar{n}_r^\text{pre}$ are set to have a good enough SNR while minimizing qubit relaxation and measurement-induced transitions. The SNR of the pre-pulse doesn't need to be maximal since we can calibrate the pre-selection thresholds to exclude data which falls within the region where the $\ket{0}$ and $\ket{1}$ gaussians overlap.

Finally the parameters of the readout pulse are tuned. The duration of the pulse $T_{\text{r}}$ is optimized to get good SNR while minimizing relaxation errors. The power of the pulse $P_{\text{r}}$ is chosen to get maximum SNR without being affected by high-power non-QND effects.

\section{Number of photon calibration}
\label{apx:nphcalib}

\begin{figure}
\centering
\includegraphics[width=1\columnwidth]{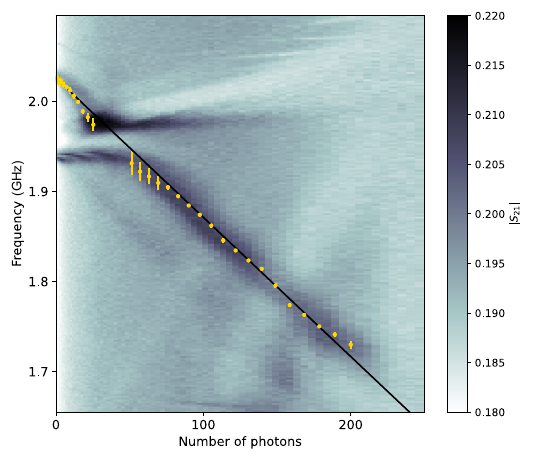}
\caption{\label{afig:nph_calib} AC-Stark shift measurement: a spectroscopy of the qubit is performed in the presence of a variable power cavity drive. The qubit's frequency is fitted for each readout power and plotted as gold points, with the fit's uncertainty. The ovelayed black line is a linear fit for the qubit's frequency shift.
}
\end{figure}

Throughout this work, the readout power is expressed as the average number of photons populating the pointer-state for $\ket{0}$, using the same convention as most works in the state-of-the-art \cite{Walter2017,Swiadek2024}. However the conversion from Watts to number of photons needs to be properly calibrated. We use an AC-Stark shift type of measurement, as was done in \cite{Schuster2005,Sank2016}, which is essentially a pulsed qubit spectroscopy performed in the presence of a readout drive of variable power and resonant with the readout mode when the qubit is in $\ket{0}$. The pulse sequence consists of a \SI{2}{\micro\second} qubit drive with a simultaneous readout drive of \SI{3}{\micro\second} at frequency $\omega_d=\omega_r$, followed by a waiting time of $9(2\pi/\kappa_r)=\SI{500}{\nano\second}$ and then by a readout pulse. The frequency of the qubit drive and the power of the readout drive are swept, giving the final result shown in Fig. \ref{afig:nph_calib}.

The qubit's frequency is tracked as a function of the readout power and shifts quite linearly as a function of power until it reaches about \SI{1.7}{\giga\hertz}, except around \SI{1.95}{\giga\hertz}, where we observe an anti-crossing that could be explained by a coupling to a parasitic TLS. This shows that the AC-Stark shift behaves linearly as a function of input readout power in the measured range, so the number of photons is extracted at each power $P$ as $\bar{n}_r(P)=\Delta\omega_q(P)/(2\chi_{qr})$, where $\Delta\omega_q(P)=\omega_q(P)-\omega_q(0)$. This linear behavior was experimentally verified up to 200 photons in Fig. \ref{afig:nph_calib} and we extrapolate it to higher power for the scale of Fig. \ref{fig:qnd_vs_params}.

\section{Lifetime and coherence results}
\label{apx:coherence}

\begin{figure}
\centering
\includegraphics[width=1\columnwidth]{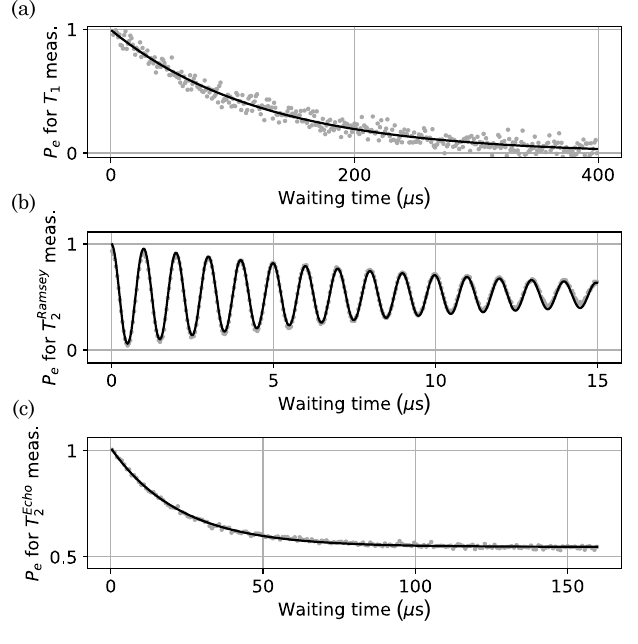}
\caption{\label{figapx:coherence} (a) Relaxation measurement yielding a fitted lifetime $T_1=\SI{124.5}{\micro\second}$. (b) Ramsey measurement, with a fitted Ramsey coherence time $T_2^*=\SI{10.6}{\micro\second}$. (c) Echo measurement, with a fitted coherence time $T_2^E=\SI{22.6}{\micro\second}$. For all plots, measured data points are shown in grey and fitted curves are shown as black lines. }
\end{figure}

The measurement results for $T_1$, $T_2^*$ and $T_2^E$ are shown in Fig. \ref{figapx:coherence}(a), Fig. \ref{figapx:coherence}(b) and Fig. \ref{figapx:coherence}(c) respectively. Compared to the results from Ref. \cite{Dassonneville2020}, the most striking change is for $T_1$, which increased from \SI{3.3}{\micro\second} to \SI{124.5}{\micro\second}. This confirms that the lifetime in our previous work was limited by remnants of Purcell effect and also shows that this effect was attenuated through this work's optimizations.
The new limitation for $T_1$ is most probably dielectric losses. A quick estimation can be done using the following formula \cite{Nguyen2022}:
\begin{equation}
    \label{eq:T1_diel}
    \Gamma_1^\text{diel}=\frac{\hbar\omega_q^2}{4E_{Cq}Q_\text{diel}}\left|\bra{0}\hat{\varphi}_q\ket{1}\right|^2\left[\coth\left(\frac{\hbar\omega_q}{2k_bT}\right)+1\right],
\end{equation}
where $\omega_q$ is the qubit's frequency, $E_{Cq}$ is its charging energy and $\hat{\varphi}_q$ is its superconducting phase operator. $Q_\text{diel}$ is the dielectric's quality factor, $k_b$ is Boltzmann's constant and $T$ is the sample's temperature. If we assume our current $T_1$ is limited by such dielectric losses, then we find an effect $Q_\text{diel}=\SI{2.67e6}{}$, which falls in the correct range for aluminium on silicon samples. This can be later improved upon by changing the materials used to make the transmon molecule. Newer samples made from tantalum on sapphire can achieve a higher $Q_\text{diel}$ and yield better lifetimes.

The value for $T_2^*$ also improved, going from \SI{3.2}{\micro\second} to \SI{10.6}{\micro\second} and Echo measurements give $T_\text{2}^E=\SI{22.6}{\micro\second}$. Since the Ramsey and Echo measurements show the same order of magnitude for $T_2^*$ and $T_2^E$, this means that the noise limiting our coherence is in the high-frequency range. The most likely limitation in this case is the dephasing induced by thermal photons in the 3D resonator.
With the formula from \cite{Bertet2005}, we compute the dephasing induced by both of the readout mode and the unused lower polariton, which is noted with the index $l$. We then get the following expression for the total dephasing induced by thermal photons:
\begin{align}
\begin{split}
    \left(T_2^\text{th}\right)^{-1} =& \frac{\bar{n}_{l,\text{th}}(\bar{n}_{l,\text{th}}+1)(2\chi_{ql})^2}{\kappa_l}  \\
    &+ \frac{\bar{n}_{r,\text{th}}(\bar{n}_{r,\text{th}}+1)(2\chi_{qr})^2}{\kappa_r}
\end{split}
\end{align}
where $\bar{n}_{l,\text{th}}$ (resp. $\bar{n}_{r,\text{th}}$) is the number of thermal photons in the lower polariton (resp. readout mode), $\kappa_l$ (resp. $\kappa_r$) and $\chi_l$ (resp. $\chi_r$) are the total losses and the dispersive shift of the lower polariton  (resp. readout mode). If we assume that our $T_2^E$ is entirely due to thermal photons, and assume that $\bar{n}_{l,\text{th}}$ and $\bar{n}_{r,\text{th}}$ both follow a Bose-Einstein distribution of identical temperature, then we find an effective temperature of \SI{38.7}{\milli\kelvin}, which is consistent with our cryostat's base temperature of \SI{30}{\milli\kelvin}. This temperature results in thermal populations $\bar{n}_{l,\text{th}}=3.4\times10^{-4}$ and $\bar{n}_{r,\text{th}}=1.2\times10^{-4}$. This suggests that our $T_2^E$ is mainly limited by thermal photons in the lower polariton. This can be mitigated by increasing the frequency of the lower polariton, which is possible by tuning the sample's parameters to have a higher frequency ancilla mode. For example, if the lower polariton was designed to be at \SI{8}{\giga\hertz}, we would get $T_2^\text{th} = \SI{158} {\micro\second}$.

\bibliography{high_fid_biblio}

\begin{thebibliography}{51}%
\makeatletter
\providecommand \@ifxundefined [1]{%
 \@ifx{#1\undefined}
}%
\providecommand \@ifnum [1]{%
 \ifnum #1\expandafter \@firstoftwo
 \else \expandafter \@secondoftwo
 \fi
}%
\providecommand \@ifx [1]{%
 \ifx #1\expandafter \@firstoftwo
 \else \expandafter \@secondoftwo
 \fi
}%
\providecommand \natexlab [1]{#1}%
\providecommand \enquote  [1]{``#1''}%
\providecommand \bibnamefont  [1]{#1}%
\providecommand \bibfnamefont [1]{#1}%
\providecommand \citenamefont [1]{#1}%
\providecommand \href@noop [0]{\@secondoftwo}%
\providecommand \href [0]{\begingroup \@sanitize@url \@href}%
\providecommand \@href[1]{\@@startlink{#1}\@@href}%
\providecommand \@@href[1]{\endgroup#1\@@endlink}%
\providecommand \@sanitize@url [0]{\catcode `\\12\catcode `\$12\catcode `\&12\catcode `\#12\catcode `\^12\catcode `\_12\catcode `\%12\relax}%
\providecommand \@@startlink[1]{}%
\providecommand \@@endlink[0]{}%
\providecommand \url  [0]{\begingroup\@sanitize@url \@url }%
\providecommand \@url [1]{\endgroup\@href {#1}{\urlprefix }}%
\providecommand \urlprefix  [0]{URL }%
\providecommand \Eprint [0]{\href }%
\providecommand \doibase [0]{https://doi.org/}%
\providecommand \selectlanguage [0]{\@gobble}%
\providecommand \bibinfo  [0]{\@secondoftwo}%
\providecommand \bibfield  [0]{\@secondoftwo}%
\providecommand \translation [1]{[#1]}%
\providecommand \BibitemOpen [0]{}%
\providecommand \bibitemStop [0]{}%
\providecommand \bibitemNoStop [0]{.\EOS\space}%
\providecommand \EOS [0]{\spacefactor3000\relax}%
\providecommand \BibitemShut  [1]{\csname bibitem#1\endcsname}%
\let\auto@bib@innerbib\@empty
\bibitem [{\citenamefont {Shor}(1995)}]{Shor1995}%
  \BibitemOpen
  \bibfield  {author} {\bibinfo {author} {\bibfnamefont {P.~W.}\ \bibnamefont {Shor}},\ }\bibfield  {title} {\bibinfo {title} {Scheme for reducing decoherence in quantum computer memory},\ }\href {https://doi.org/10.1103/PhysRevA.52.R2493} {\bibfield  {journal} {\bibinfo  {journal} {Physical Review A}\ }\textbf {\bibinfo {volume} {52}},\ \bibinfo {pages} {R2493} (\bibinfo {year} {1995})}\BibitemShut {NoStop}%
\bibitem [{\citenamefont {Gottesman}(1997)}]{Gottesman1997}%
  \BibitemOpen
  \bibfield  {author} {\bibinfo {author} {\bibfnamefont {D.~E.}\ \bibnamefont {Gottesman}},\ }\emph {\bibinfo {title} {Stabilizer {Codes} and {Quantum} {Error} {Correction}}},\ \href {https://doi.org/10.7907/rzr7-dt72} {Ph.D. thesis},\ \bibinfo  {school} {California Institute of Technology} (\bibinfo {year} {1997})\BibitemShut {NoStop}%
\bibitem [{\citenamefont {Johnson}\ \emph {et~al.}(2012)\citenamefont {Johnson}, \citenamefont {Macklin}, \citenamefont {Slichter}, \citenamefont {Vijay}, \citenamefont {Weingarten}, \citenamefont {Clarke},\ and\ \citenamefont {Siddiqi}}]{Johnson2012}%
  \BibitemOpen
  \bibfield  {author} {\bibinfo {author} {\bibfnamefont {J.~E.}\ \bibnamefont {Johnson}}, \bibinfo {author} {\bibfnamefont {C.}~\bibnamefont {Macklin}}, \bibinfo {author} {\bibfnamefont {D.~H.}\ \bibnamefont {Slichter}}, \bibinfo {author} {\bibfnamefont {R.}~\bibnamefont {Vijay}}, \bibinfo {author} {\bibfnamefont {E.~B.}\ \bibnamefont {Weingarten}}, \bibinfo {author} {\bibfnamefont {J.}~\bibnamefont {Clarke}},\ and\ \bibinfo {author} {\bibfnamefont {I.}~\bibnamefont {Siddiqi}},\ }\bibfield  {title} {\bibinfo {title} {Heralded {State} {Preparation} in a {Superconducting} {Qubit}},\ }\href {https://doi.org/10.1103/PhysRevLett.109.050506} {\bibfield  {journal} {\bibinfo  {journal} {Physical Review Letters}\ }\textbf {\bibinfo {volume} {109}},\ \bibinfo {pages} {050506} (\bibinfo {year} {2012})}\BibitemShut {NoStop}%
\bibitem [{\citenamefont {Ristè}\ \emph {et~al.}(2012)\citenamefont {Ristè}, \citenamefont {Bultink}, \citenamefont {Lehnert},\ and\ \citenamefont {DiCarlo}}]{Riste2012}%
  \BibitemOpen
  \bibfield  {author} {\bibinfo {author} {\bibfnamefont {D.}~\bibnamefont {Ristè}}, \bibinfo {author} {\bibfnamefont {C.~C.}\ \bibnamefont {Bultink}}, \bibinfo {author} {\bibfnamefont {K.~W.}\ \bibnamefont {Lehnert}},\ and\ \bibinfo {author} {\bibfnamefont {L.}~\bibnamefont {DiCarlo}},\ }\bibfield  {title} {\bibinfo {title} {Feedback {Control} of a {Solid}-{State} {Qubit} {Using} {High}-{Fidelity} {Projective} {Measurement}},\ }\href {https://doi.org/10.1103/PhysRevLett.109.240502} {\bibfield  {journal} {\bibinfo  {journal} {Physical Review Letters}\ }\textbf {\bibinfo {volume} {109}},\ \bibinfo {pages} {240502} (\bibinfo {year} {2012})}\BibitemShut {NoStop}%
\bibitem [{\citenamefont {Koch}\ \emph {et~al.}(2007)\citenamefont {Koch}, \citenamefont {Yu}, \citenamefont {Gambetta}, \citenamefont {Houck}, \citenamefont {Schuster}, \citenamefont {Majer}, \citenamefont {Blais}, \citenamefont {Devoret}, \citenamefont {Girvin},\ and\ \citenamefont {Schoelkopf}}]{Koch2007}%
  \BibitemOpen
  \bibfield  {author} {\bibinfo {author} {\bibfnamefont {J.}~\bibnamefont {Koch}}, \bibinfo {author} {\bibfnamefont {T.~M.}\ \bibnamefont {Yu}}, \bibinfo {author} {\bibfnamefont {J.}~\bibnamefont {Gambetta}}, \bibinfo {author} {\bibfnamefont {A.~A.}\ \bibnamefont {Houck}}, \bibinfo {author} {\bibfnamefont {D.~I.}\ \bibnamefont {Schuster}}, \bibinfo {author} {\bibfnamefont {J.}~\bibnamefont {Majer}}, \bibinfo {author} {\bibfnamefont {A.}~\bibnamefont {Blais}}, \bibinfo {author} {\bibfnamefont {M.~H.}\ \bibnamefont {Devoret}}, \bibinfo {author} {\bibfnamefont {S.~M.}\ \bibnamefont {Girvin}},\ and\ \bibinfo {author} {\bibfnamefont {R.~J.}\ \bibnamefont {Schoelkopf}},\ }\bibfield  {title} {\bibinfo {title} {Charge-insensitive qubit design derived from the {Cooper} pair box},\ }\href {https://doi.org/10.1103/PhysRevA.76.042319} {\bibfield  {journal} {\bibinfo  {journal} {Physical Review A}\ }\textbf {\bibinfo {volume} {76}},\ \bibinfo {pages} {042319} (\bibinfo {year} {2007})}\BibitemShut {NoStop}%
\bibitem [{\citenamefont {Swiadek}\ \emph {et~al.}(2024)\citenamefont {Swiadek}, \citenamefont {Shillito}, \citenamefont {Magnard}, \citenamefont {Remm}, \citenamefont {Hellings}, \citenamefont {Lacroix}, \citenamefont {Ficheux}, \citenamefont {Zanuz}, \citenamefont {Norris}, \citenamefont {Blais}, \citenamefont {Krinner},\ and\ \citenamefont {Wallraff}}]{Swiadek2024}%
  \BibitemOpen
  \bibfield  {author} {\bibinfo {author} {\bibfnamefont {F.}~\bibnamefont {Swiadek}}, \bibinfo {author} {\bibfnamefont {R.}~\bibnamefont {Shillito}}, \bibinfo {author} {\bibfnamefont {P.}~\bibnamefont {Magnard}}, \bibinfo {author} {\bibfnamefont {A.}~\bibnamefont {Remm}}, \bibinfo {author} {\bibfnamefont {C.}~\bibnamefont {Hellings}}, \bibinfo {author} {\bibfnamefont {N.}~\bibnamefont {Lacroix}}, \bibinfo {author} {\bibfnamefont {Q.}~\bibnamefont {Ficheux}}, \bibinfo {author} {\bibfnamefont {D.~C.}\ \bibnamefont {Zanuz}}, \bibinfo {author} {\bibfnamefont {G.~J.}\ \bibnamefont {Norris}}, \bibinfo {author} {\bibfnamefont {A.}~\bibnamefont {Blais}}, \bibinfo {author} {\bibfnamefont {S.}~\bibnamefont {Krinner}},\ and\ \bibinfo {author} {\bibfnamefont {A.}~\bibnamefont {Wallraff}},\ }\bibfield  {title} {\bibinfo {title} {Enhancing {Dispersive} {Readout} of {Superconducting} {Qubits} through {Dynamic} {Control} of the {Dispersive} {Shift}: {Experiment} and {Theory}},\ }\href
  {https://doi.org/10.1103/PRXQuantum.5.040326} {\bibfield  {journal} {\bibinfo  {journal} {PRX Quantum}\ }\textbf {\bibinfo {volume} {5}},\ \bibinfo {pages} {040326} (\bibinfo {year} {2024})}\BibitemShut {NoStop}%
\bibitem [{\citenamefont {Kurilovich}\ \emph {et~al.}(2025)\citenamefont {Kurilovich}, \citenamefont {Connolly}, \citenamefont {Bøttcher}, \citenamefont {Weiss}, \citenamefont {Hazra}, \citenamefont {Joshi}, \citenamefont {Ding}, \citenamefont {Nho}, \citenamefont {Diamond}, \citenamefont {Kurilovich}, \citenamefont {Dai}, \citenamefont {Fatemi}, \citenamefont {Frunzio}, \citenamefont {Glazman},\ and\ \citenamefont {Devoret}}]{Kurilovich2025}%
  \BibitemOpen
  \bibfield  {author} {\bibinfo {author} {\bibfnamefont {P.~D.}\ \bibnamefont {Kurilovich}}, \bibinfo {author} {\bibfnamefont {T.}~\bibnamefont {Connolly}}, \bibinfo {author} {\bibfnamefont {C.~G.~L.}\ \bibnamefont {Bøttcher}}, \bibinfo {author} {\bibfnamefont {D.~K.}\ \bibnamefont {Weiss}}, \bibinfo {author} {\bibfnamefont {S.}~\bibnamefont {Hazra}}, \bibinfo {author} {\bibfnamefont {V.~R.}\ \bibnamefont {Joshi}}, \bibinfo {author} {\bibfnamefont {A.~Z.}\ \bibnamefont {Ding}}, \bibinfo {author} {\bibfnamefont {H.}~\bibnamefont {Nho}}, \bibinfo {author} {\bibfnamefont {S.}~\bibnamefont {Diamond}}, \bibinfo {author} {\bibfnamefont {V.~D.}\ \bibnamefont {Kurilovich}}, \bibinfo {author} {\bibfnamefont {W.}~\bibnamefont {Dai}}, \bibinfo {author} {\bibfnamefont {V.}~\bibnamefont {Fatemi}}, \bibinfo {author} {\bibfnamefont {L.}~\bibnamefont {Frunzio}}, \bibinfo {author} {\bibfnamefont {L.~I.}\ \bibnamefont {Glazman}},\ and\ \bibinfo {author} {\bibfnamefont {M.~H.}\ \bibnamefont {Devoret}},\ }\bibfield  {title}
  {\bibinfo {title} {High-frequency readout free from transmon multi-excitation resonances}} (\bibinfo {year} {2025}),\ \bibinfo {note} {arXiv:2501.09161 [quant-ph]}\BibitemShut {NoStop}%
\bibitem [{\citenamefont {Connolly}\ \emph {et~al.}(2025)\citenamefont {Connolly}, \citenamefont {Kurilovich}, \citenamefont {Kurilovich}, \citenamefont {Bøttcher}, \citenamefont {Hazra}, \citenamefont {Dai}, \citenamefont {Ding}, \citenamefont {Joshi}, \citenamefont {Nho}, \citenamefont {Diamond}, \citenamefont {Weiss}, \citenamefont {Fatemi}, \citenamefont {Frunzio}, \citenamefont {Glazman},\ and\ \citenamefont {Devoret}}]{Connolly2025}%
  \BibitemOpen
  \bibfield  {author} {\bibinfo {author} {\bibfnamefont {T.}~\bibnamefont {Connolly}}, \bibinfo {author} {\bibfnamefont {P.~D.}\ \bibnamefont {Kurilovich}}, \bibinfo {author} {\bibfnamefont {V.~D.}\ \bibnamefont {Kurilovich}}, \bibinfo {author} {\bibfnamefont {C.~G.~L.}\ \bibnamefont {Bøttcher}}, \bibinfo {author} {\bibfnamefont {S.}~\bibnamefont {Hazra}}, \bibinfo {author} {\bibfnamefont {W.}~\bibnamefont {Dai}}, \bibinfo {author} {\bibfnamefont {A.~Z.}\ \bibnamefont {Ding}}, \bibinfo {author} {\bibfnamefont {V.~R.}\ \bibnamefont {Joshi}}, \bibinfo {author} {\bibfnamefont {H.}~\bibnamefont {Nho}}, \bibinfo {author} {\bibfnamefont {S.}~\bibnamefont {Diamond}}, \bibinfo {author} {\bibfnamefont {D.~K.}\ \bibnamefont {Weiss}}, \bibinfo {author} {\bibfnamefont {V.}~\bibnamefont {Fatemi}}, \bibinfo {author} {\bibfnamefont {L.}~\bibnamefont {Frunzio}}, \bibinfo {author} {\bibfnamefont {L.~I.}\ \bibnamefont {Glazman}},\ and\ \bibinfo {author} {\bibfnamefont {M.~H.}\ \bibnamefont {Devoret}},\ }\bibfield  {title}
  {\bibinfo {title} {Full characterization of measurement-induced transitions of a superconducting qubit}} (\bibinfo {year} {2025}),\ \bibinfo {note} {arXiv:2506.05306 [quant-ph]}\BibitemShut {NoStop}%
\bibitem [{\citenamefont {Gusenkova}\ \emph {et~al.}(2021)\citenamefont {Gusenkova}, \citenamefont {Spiecker}, \citenamefont {Gebauer}, \citenamefont {Willsch}, \citenamefont {Willsch}, \citenamefont {Valenti}, \citenamefont {Karcher}, \citenamefont {Grünhaupt}, \citenamefont {Takmakov}, \citenamefont {Winkel}, \citenamefont {Rieger}, \citenamefont {Ustinov}, \citenamefont {Roch}, \citenamefont {Wernsdorfer}, \citenamefont {Michielsen}, \citenamefont {Sander},\ and\ \citenamefont {Pop}}]{Gusenkova2021}%
  \BibitemOpen
  \bibfield  {author} {\bibinfo {author} {\bibfnamefont {D.}~\bibnamefont {Gusenkova}}, \bibinfo {author} {\bibfnamefont {M.}~\bibnamefont {Spiecker}}, \bibinfo {author} {\bibfnamefont {R.}~\bibnamefont {Gebauer}}, \bibinfo {author} {\bibfnamefont {M.}~\bibnamefont {Willsch}}, \bibinfo {author} {\bibfnamefont {D.}~\bibnamefont {Willsch}}, \bibinfo {author} {\bibfnamefont {F.}~\bibnamefont {Valenti}}, \bibinfo {author} {\bibfnamefont {N.}~\bibnamefont {Karcher}}, \bibinfo {author} {\bibfnamefont {L.}~\bibnamefont {Grünhaupt}}, \bibinfo {author} {\bibfnamefont {I.}~\bibnamefont {Takmakov}}, \bibinfo {author} {\bibfnamefont {P.}~\bibnamefont {Winkel}}, \bibinfo {author} {\bibfnamefont {D.}~\bibnamefont {Rieger}}, \bibinfo {author} {\bibfnamefont {A.~V.}\ \bibnamefont {Ustinov}}, \bibinfo {author} {\bibfnamefont {N.}~\bibnamefont {Roch}}, \bibinfo {author} {\bibfnamefont {W.}~\bibnamefont {Wernsdorfer}}, \bibinfo {author} {\bibfnamefont {K.}~\bibnamefont {Michielsen}}, \bibinfo {author} {\bibfnamefont
  {O.}~\bibnamefont {Sander}},\ and\ \bibinfo {author} {\bibfnamefont {I.~M.}\ \bibnamefont {Pop}},\ }\bibfield  {title} {\bibinfo {title} {Quantum {Nondemolition} {Dispersive} {Readout} of a {Superconducting} {Artificial} {Atom} {Using} {Large} {Photon} {Numbers}},\ }\href {https://doi.org/10.1103/PhysRevApplied.15.064030} {\bibfield  {journal} {\bibinfo  {journal} {Physical Review Applied}\ }\textbf {\bibinfo {volume} {15}},\ \bibinfo {pages} {064030} (\bibinfo {year} {2021})}\BibitemShut {NoStop}%
\bibitem [{\citenamefont {Takmakov}\ \emph {et~al.}(2021)\citenamefont {Takmakov}, \citenamefont {Winkel}, \citenamefont {Foroughi}, \citenamefont {Planat}, \citenamefont {Gusenkova}, \citenamefont {Spiecker}, \citenamefont {Rieger}, \citenamefont {Grünhaupt}, \citenamefont {Ustinov}, \citenamefont {Wernsdorfer}, \citenamefont {Pop},\ and\ \citenamefont {Roch}}]{Takmakov2021}%
  \BibitemOpen
  \bibfield  {author} {\bibinfo {author} {\bibfnamefont {I.}~\bibnamefont {Takmakov}}, \bibinfo {author} {\bibfnamefont {P.}~\bibnamefont {Winkel}}, \bibinfo {author} {\bibfnamefont {F.}~\bibnamefont {Foroughi}}, \bibinfo {author} {\bibfnamefont {L.}~\bibnamefont {Planat}}, \bibinfo {author} {\bibfnamefont {D.}~\bibnamefont {Gusenkova}}, \bibinfo {author} {\bibfnamefont {M.}~\bibnamefont {Spiecker}}, \bibinfo {author} {\bibfnamefont {D.}~\bibnamefont {Rieger}}, \bibinfo {author} {\bibfnamefont {L.}~\bibnamefont {Grünhaupt}}, \bibinfo {author} {\bibfnamefont {A.}~\bibnamefont {Ustinov}}, \bibinfo {author} {\bibfnamefont {W.}~\bibnamefont {Wernsdorfer}}, \bibinfo {author} {\bibfnamefont {I.}~\bibnamefont {Pop}},\ and\ \bibinfo {author} {\bibfnamefont {N.}~\bibnamefont {Roch}},\ }\bibfield  {title} {\bibinfo {title} {Minimizing the {Discrimination} {Time} for {Quantum} {States} of an {Artificial} {Atom}},\ }\href {https://doi.org/10.1103/PhysRevApplied.15.064029} {\bibfield  {journal} {\bibinfo  {journal}
  {Physical Review Applied}\ }\textbf {\bibinfo {volume} {15}},\ \bibinfo {pages} {064029} (\bibinfo {year} {2021})}\BibitemShut {NoStop}%
\bibitem [{\citenamefont {Diniz}\ \emph {et~al.}(2013)\citenamefont {Diniz}, \citenamefont {Dumur}, \citenamefont {Buisson},\ and\ \citenamefont {Auffèves}}]{Diniz2013}%
  \BibitemOpen
  \bibfield  {author} {\bibinfo {author} {\bibfnamefont {I.}~\bibnamefont {Diniz}}, \bibinfo {author} {\bibfnamefont {E.}~\bibnamefont {Dumur}}, \bibinfo {author} {\bibfnamefont {O.}~\bibnamefont {Buisson}},\ and\ \bibinfo {author} {\bibfnamefont {A.}~\bibnamefont {Auffèves}},\ }\bibfield  {title} {\bibinfo {title} {Ultrafast quantum nondemolition measurements based on a diamond-shaped artificial atom},\ }\href {https://doi.org/10.1103/PhysRevA.87.033837} {\bibfield  {journal} {\bibinfo  {journal} {Physical Review A}\ }\textbf {\bibinfo {volume} {87}},\ \bibinfo {pages} {033837} (\bibinfo {year} {2013})}\BibitemShut {NoStop}%
\bibitem [{\citenamefont {Dassonneville}\ \emph {et~al.}(2020)\citenamefont {Dassonneville}, \citenamefont {Ramos}, \citenamefont {Milchakov}, \citenamefont {Planat}, \citenamefont {Dumur}, \citenamefont {Foroughi}, \citenamefont {Puertas}, \citenamefont {Leger}, \citenamefont {Bharadwaj}, \citenamefont {Delaforce}, \citenamefont {Naud}, \citenamefont {Hasch-Guichard}, \citenamefont {García-Ripoll}, \citenamefont {Roch},\ and\ \citenamefont {Buisson}}]{Dassonneville2020}%
  \BibitemOpen
  \bibfield  {author} {\bibinfo {author} {\bibfnamefont {R.}~\bibnamefont {Dassonneville}}, \bibinfo {author} {\bibfnamefont {T.}~\bibnamefont {Ramos}}, \bibinfo {author} {\bibfnamefont {V.}~\bibnamefont {Milchakov}}, \bibinfo {author} {\bibfnamefont {L.}~\bibnamefont {Planat}}, \bibinfo {author} {\bibfnamefont {E.}~\bibnamefont {Dumur}}, \bibinfo {author} {\bibfnamefont {F.}~\bibnamefont {Foroughi}}, \bibinfo {author} {\bibfnamefont {J.}~\bibnamefont {Puertas}}, \bibinfo {author} {\bibfnamefont {S.}~\bibnamefont {Leger}}, \bibinfo {author} {\bibfnamefont {K.}~\bibnamefont {Bharadwaj}}, \bibinfo {author} {\bibfnamefont {J.}~\bibnamefont {Delaforce}}, \bibinfo {author} {\bibfnamefont {C.}~\bibnamefont {Naud}}, \bibinfo {author} {\bibfnamefont {W.}~\bibnamefont {Hasch-Guichard}}, \bibinfo {author} {\bibfnamefont {J.}~\bibnamefont {García-Ripoll}}, \bibinfo {author} {\bibfnamefont {N.}~\bibnamefont {Roch}},\ and\ \bibinfo {author} {\bibfnamefont {O.}~\bibnamefont {Buisson}},\ }\bibfield  {title} {\bibinfo
  {title} {Fast {High}-{Fidelity} {Quantum} {Nondemolition} {Qubit} {Readout} via a {Nonperturbative} {Cross}-{Kerr} {Coupling}},\ }\href {https://doi.org/10.1103/PhysRevX.10.011045} {\bibfield  {journal} {\bibinfo  {journal} {Physical Review X}\ }\textbf {\bibinfo {volume} {10}},\ \bibinfo {pages} {011045} (\bibinfo {year} {2020})}\BibitemShut {NoStop}%
\bibitem [{\citenamefont {Ye}\ \emph {et~al.}(2024)\citenamefont {Ye}, \citenamefont {Kline}, \citenamefont {Chen}, \citenamefont {Yen},\ and\ \citenamefont {O’Brien}}]{Ye2024}%
  \BibitemOpen
  \bibfield  {author} {\bibinfo {author} {\bibfnamefont {Y.}~\bibnamefont {Ye}}, \bibinfo {author} {\bibfnamefont {J.~B.}\ \bibnamefont {Kline}}, \bibinfo {author} {\bibfnamefont {S.}~\bibnamefont {Chen}}, \bibinfo {author} {\bibfnamefont {A.}~\bibnamefont {Yen}},\ and\ \bibinfo {author} {\bibfnamefont {K.~P.}\ \bibnamefont {O’Brien}},\ }\bibfield  {title} {\bibinfo {title} {Ultrafast superconducting qubit readout with the quarton coupler},\ }\bibfield  {journal} {\bibinfo  {journal} {Science Advances}\ }\textbf {\bibinfo {volume} {10}},\ \href {https://doi.org/10.1126/sciadv.ado9094} {10.1126/sciadv.ado9094} (\bibinfo {year} {2024})\BibitemShut {NoStop}%
\bibitem [{\citenamefont {Pfeiffer}\ \emph {et~al.}(2024)\citenamefont {Pfeiffer}, \citenamefont {Werninghaus}, \citenamefont {Schweizer}, \citenamefont {Bruckmoser}, \citenamefont {Koch}, \citenamefont {Glaser}, \citenamefont {Huber}, \citenamefont {Bunch}, \citenamefont {Haslbeck}, \citenamefont {Knudsen}, \citenamefont {Krylov}, \citenamefont {Liegener}, \citenamefont {Marx}, \citenamefont {Richard}, \citenamefont {Romeiro}, \citenamefont {Roy}, \citenamefont {Schirk}, \citenamefont {Schneider}, \citenamefont {Singh}, \citenamefont {Södergren}, \citenamefont {Tsitsilin}, \citenamefont {Wallner}, \citenamefont {Riofrío},\ and\ \citenamefont {Filipp}}]{Pfeiffer2024}%
  \BibitemOpen
  \bibfield  {author} {\bibinfo {author} {\bibfnamefont {F.}~\bibnamefont {Pfeiffer}}, \bibinfo {author} {\bibfnamefont {M.}~\bibnamefont {Werninghaus}}, \bibinfo {author} {\bibfnamefont {C.}~\bibnamefont {Schweizer}}, \bibinfo {author} {\bibfnamefont {N.}~\bibnamefont {Bruckmoser}}, \bibinfo {author} {\bibfnamefont {L.}~\bibnamefont {Koch}}, \bibinfo {author} {\bibfnamefont {N.}~\bibnamefont {Glaser}}, \bibinfo {author} {\bibfnamefont {G.}~\bibnamefont {Huber}}, \bibinfo {author} {\bibfnamefont {D.}~\bibnamefont {Bunch}}, \bibinfo {author} {\bibfnamefont {F.}~\bibnamefont {Haslbeck}}, \bibinfo {author} {\bibfnamefont {M.}~\bibnamefont {Knudsen}}, \bibinfo {author} {\bibfnamefont {G.}~\bibnamefont {Krylov}}, \bibinfo {author} {\bibfnamefont {K.}~\bibnamefont {Liegener}}, \bibinfo {author} {\bibfnamefont {A.}~\bibnamefont {Marx}}, \bibinfo {author} {\bibfnamefont {L.}~\bibnamefont {Richard}}, \bibinfo {author} {\bibfnamefont {J.}~\bibnamefont {Romeiro}}, \bibinfo {author} {\bibfnamefont {F.}~\bibnamefont
  {Roy}}, \bibinfo {author} {\bibfnamefont {J.}~\bibnamefont {Schirk}}, \bibinfo {author} {\bibfnamefont {C.}~\bibnamefont {Schneider}}, \bibinfo {author} {\bibfnamefont {M.}~\bibnamefont {Singh}}, \bibinfo {author} {\bibfnamefont {L.}~\bibnamefont {Södergren}}, \bibinfo {author} {\bibfnamefont {I.}~\bibnamefont {Tsitsilin}}, \bibinfo {author} {\bibfnamefont {F.}~\bibnamefont {Wallner}}, \bibinfo {author} {\bibfnamefont {C.}~\bibnamefont {Riofrío}},\ and\ \bibinfo {author} {\bibfnamefont {S.}~\bibnamefont {Filipp}},\ }\bibfield  {title} {\bibinfo {title} {Efficient {Decoupling} of a {Nonlinear} {Qubit} {Mode} from {Its} {Environment}},\ }\href {https://doi.org/10.1103/PhysRevX.14.041007} {\bibfield  {journal} {\bibinfo  {journal} {Physical Review X}\ }\textbf {\bibinfo {volume} {14}},\ \bibinfo {pages} {041007} (\bibinfo {year} {2024})}\BibitemShut {NoStop}%
\bibitem [{\citenamefont {Salunkhe}\ \emph {et~al.}(2025)\citenamefont {Salunkhe}, \citenamefont {Kundu}, \citenamefont {Das}, \citenamefont {Deshmukh}, \citenamefont {Patankar},\ and\ \citenamefont {Vijay}}]{Salunkhe2025}%
  \BibitemOpen
  \bibfield  {author} {\bibinfo {author} {\bibfnamefont {K.~V.}\ \bibnamefont {Salunkhe}}, \bibinfo {author} {\bibfnamefont {S.}~\bibnamefont {Kundu}}, \bibinfo {author} {\bibfnamefont {S.}~\bibnamefont {Das}}, \bibinfo {author} {\bibfnamefont {J.}~\bibnamefont {Deshmukh}}, \bibinfo {author} {\bibfnamefont {M.~P.}\ \bibnamefont {Patankar}},\ and\ \bibinfo {author} {\bibfnamefont {R.}~\bibnamefont {Vijay}},\ }\bibfield  {title} {\bibinfo {title} {The quantromon: A qubit-resonator system with orthogonal qubit and readout modes},\ }\Eprint {https://arxiv.org/abs/2501.17439} {arXiv:2501.17439 [quant-ph]}  (\bibinfo {year} {2025})\BibitemShut {NoStop}%
\bibitem [{\citenamefont {Dassonneville}\ \emph {et~al.}(2023)\citenamefont {Dassonneville}, \citenamefont {Ramos}, \citenamefont {Milchakov}, \citenamefont {Mori}, \citenamefont {Planat}, \citenamefont {Foroughi}, \citenamefont {Naud}, \citenamefont {Hasch-Guichard}, \citenamefont {García-Ripoll}, \citenamefont {Roch},\ and\ \citenamefont {Buisson}}]{Dassonneville2023}%
  \BibitemOpen
  \bibfield  {author} {\bibinfo {author} {\bibfnamefont {R.}~\bibnamefont {Dassonneville}}, \bibinfo {author} {\bibfnamefont {T.}~\bibnamefont {Ramos}}, \bibinfo {author} {\bibfnamefont {V.}~\bibnamefont {Milchakov}}, \bibinfo {author} {\bibfnamefont {C.}~\bibnamefont {Mori}}, \bibinfo {author} {\bibfnamefont {L.}~\bibnamefont {Planat}}, \bibinfo {author} {\bibfnamefont {F.}~\bibnamefont {Foroughi}}, \bibinfo {author} {\bibfnamefont {C.}~\bibnamefont {Naud}}, \bibinfo {author} {\bibfnamefont {W.}~\bibnamefont {Hasch-Guichard}}, \bibinfo {author} {\bibfnamefont {J.}~\bibnamefont {García-Ripoll}}, \bibinfo {author} {\bibfnamefont {N.}~\bibnamefont {Roch}},\ and\ \bibinfo {author} {\bibfnamefont {O.}~\bibnamefont {Buisson}},\ }\bibfield  {title} {\bibinfo {title} {Transmon-qubit readout using an in situ bifurcation amplification in the mesoscopic regime},\ }\href {https://doi.org/10.1103/PhysRevApplied.20.044050} {\bibfield  {journal} {\bibinfo  {journal} {Physical Review Applied}\ }\textbf {\bibinfo {volume}
  {20}},\ \bibinfo {pages} {044050} (\bibinfo {year} {2023})}\BibitemShut {NoStop}%
\bibitem [{\citenamefont {Didier}\ \emph {et~al.}(2015)\citenamefont {Didier}, \citenamefont {Bourassa},\ and\ \citenamefont {Blais}}]{Didier2015}%
  \BibitemOpen
  \bibfield  {author} {\bibinfo {author} {\bibfnamefont {N.}~\bibnamefont {Didier}}, \bibinfo {author} {\bibfnamefont {J.}~\bibnamefont {Bourassa}},\ and\ \bibinfo {author} {\bibfnamefont {A.}~\bibnamefont {Blais}},\ }\bibfield  {title} {\bibinfo {title} {Fast {Quantum} {Nondemolition} {Readout} by {Parametric} {Modulation} of {Longitudinal} {Qubit}-{Oscillator} {Interaction}},\ }\href {https://doi.org/10.1103/PhysRevLett.115.203601} {\bibfield  {journal} {\bibinfo  {journal} {Physical Review Letters}\ }\textbf {\bibinfo {volume} {115}},\ \bibinfo {pages} {203601} (\bibinfo {year} {2015})}\BibitemShut {NoStop}%
\bibitem [{\citenamefont {Chapple}\ \emph {et~al.}(2024)\citenamefont {Chapple}, \citenamefont {McDonald}, \citenamefont {Muñoz-Arias},\ and\ \citenamefont {Blais}}]{Chapple2024}%
  \BibitemOpen
  \bibfield  {author} {\bibinfo {author} {\bibfnamefont {A.~A.}\ \bibnamefont {Chapple}}, \bibinfo {author} {\bibfnamefont {A.}~\bibnamefont {McDonald}}, \bibinfo {author} {\bibfnamefont {M.~H.}\ \bibnamefont {Muñoz-Arias}},\ and\ \bibinfo {author} {\bibfnamefont {A.}~\bibnamefont {Blais}},\ }\bibfield  {title} {\bibinfo {title} {Robustness of longitudinal transmon readout to ionization}} (\bibinfo {year} {2024}),\ \bibinfo {note} {arXiv:2412.07734 [quant-ph]}\BibitemShut {NoStop}%
\bibitem [{\citenamefont {Blais}\ \emph {et~al.}(2004)\citenamefont {Blais}, \citenamefont {Huang}, \citenamefont {Wallraff}, \citenamefont {Girvin},\ and\ \citenamefont {Schoelkopf}}]{Blais2004}%
  \BibitemOpen
  \bibfield  {author} {\bibinfo {author} {\bibfnamefont {A.}~\bibnamefont {Blais}}, \bibinfo {author} {\bibfnamefont {R.-S.}\ \bibnamefont {Huang}}, \bibinfo {author} {\bibfnamefont {A.}~\bibnamefont {Wallraff}}, \bibinfo {author} {\bibfnamefont {S.~M.}\ \bibnamefont {Girvin}},\ and\ \bibinfo {author} {\bibfnamefont {R.~J.}\ \bibnamefont {Schoelkopf}},\ }\bibfield  {title} {\bibinfo {title} {Cavity quantum electrodynamics for superconducting electrical circuits: {An} architecture for quantum computation},\ }\href {https://doi.org/10.1103/PhysRevA.69.062320} {\bibfield  {journal} {\bibinfo  {journal} {Physical Review A}\ }\textbf {\bibinfo {volume} {69}},\ \bibinfo {pages} {062320} (\bibinfo {year} {2004})}\BibitemShut {NoStop}%
\bibitem [{\citenamefont {Gambetta}\ \emph {et~al.}(2006)\citenamefont {Gambetta}, \citenamefont {Blais}, \citenamefont {Schuster}, \citenamefont {Wallraff}, \citenamefont {Frunzio}, \citenamefont {Majer}, \citenamefont {Devoret}, \citenamefont {Girvin},\ and\ \citenamefont {Schoelkopf}}]{Gambetta_2006}%
  \BibitemOpen
  \bibfield  {author} {\bibinfo {author} {\bibfnamefont {J.}~\bibnamefont {Gambetta}}, \bibinfo {author} {\bibfnamefont {A.}~\bibnamefont {Blais}}, \bibinfo {author} {\bibfnamefont {D.~I.}\ \bibnamefont {Schuster}}, \bibinfo {author} {\bibfnamefont {A.}~\bibnamefont {Wallraff}}, \bibinfo {author} {\bibfnamefont {L.}~\bibnamefont {Frunzio}}, \bibinfo {author} {\bibfnamefont {J.}~\bibnamefont {Majer}}, \bibinfo {author} {\bibfnamefont {M.~H.}\ \bibnamefont {Devoret}}, \bibinfo {author} {\bibfnamefont {S.~M.}\ \bibnamefont {Girvin}},\ and\ \bibinfo {author} {\bibfnamefont {R.~J.}\ \bibnamefont {Schoelkopf}},\ }\bibfield  {title} {\bibinfo {title} {Qubit-photon interactions in a cavity: {Measurement}-induced dephasing and number splitting},\ }\href {https://doi.org/10.1103/PhysRevA.74.042318} {\bibfield  {journal} {\bibinfo  {journal} {Physical Review A}\ }\textbf {\bibinfo {volume} {74}},\ \bibinfo {pages} {042318} (\bibinfo {year} {2006})}\BibitemShut {NoStop}%
\bibitem [{\citenamefont {Dumur}\ \emph {et~al.}(2016)\citenamefont {Dumur}, \citenamefont {Küng}, \citenamefont {Feofanov}, \citenamefont {Weißl}, \citenamefont {Krupko}, \citenamefont {Roch}, \citenamefont {Naud}, \citenamefont {Guichard},\ and\ \citenamefont {Buisson}}]{Dumur2016}%
  \BibitemOpen
  \bibfield  {author} {\bibinfo {author} {\bibfnamefont {E.}~\bibnamefont {Dumur}}, \bibinfo {author} {\bibfnamefont {B.}~\bibnamefont {Küng}}, \bibinfo {author} {\bibfnamefont {A.}~\bibnamefont {Feofanov}}, \bibinfo {author} {\bibfnamefont {T.}~\bibnamefont {Weißl}}, \bibinfo {author} {\bibfnamefont {Y.}~\bibnamefont {Krupko}}, \bibinfo {author} {\bibfnamefont {N.}~\bibnamefont {Roch}}, \bibinfo {author} {\bibfnamefont {C.}~\bibnamefont {Naud}}, \bibinfo {author} {\bibfnamefont {W.}~\bibnamefont {Guichard}},\ and\ \bibinfo {author} {\bibfnamefont {O.}~\bibnamefont {Buisson}},\ }\bibfield  {title} {\bibinfo {title} {Unexpectedly {Allowed} {Transition} in {Two} {Inductively} {Coupled} {Transmons}},\ }\href {https://doi.org/10.1109/TASC.2016.2515020} {\bibfield  {journal} {\bibinfo  {journal} {IEEE Transactions on Applied Superconductivity}\ }\textbf {\bibinfo {volume} {26}},\ \bibinfo {pages} {1} (\bibinfo {year} {2016})}\BibitemShut {NoStop}%
\bibitem [{\citenamefont {Wallraff}\ \emph {et~al.}(2005)\citenamefont {Wallraff}, \citenamefont {Schuster}, \citenamefont {Blais}, \citenamefont {Frunzio}, \citenamefont {Majer}, \citenamefont {Devoret}, \citenamefont {Girvin},\ and\ \citenamefont {Schoelkopf}}]{Wallraff2005}%
  \BibitemOpen
  \bibfield  {author} {\bibinfo {author} {\bibfnamefont {A.}~\bibnamefont {Wallraff}}, \bibinfo {author} {\bibfnamefont {D.~I.}\ \bibnamefont {Schuster}}, \bibinfo {author} {\bibfnamefont {A.}~\bibnamefont {Blais}}, \bibinfo {author} {\bibfnamefont {L.}~\bibnamefont {Frunzio}}, \bibinfo {author} {\bibfnamefont {J.}~\bibnamefont {Majer}}, \bibinfo {author} {\bibfnamefont {M.~H.}\ \bibnamefont {Devoret}}, \bibinfo {author} {\bibfnamefont {S.~M.}\ \bibnamefont {Girvin}},\ and\ \bibinfo {author} {\bibfnamefont {R.~J.}\ \bibnamefont {Schoelkopf}},\ }\bibfield  {title} {\bibinfo {title} {Approaching {Unit} {Visibility} for {Control} of a {Superconducting} {Qubit} with {Dispersive} {Readout}},\ }\href {https://doi.org/10.1103/PhysRevLett.95.060501} {\bibfield  {journal} {\bibinfo  {journal} {Physical Review Letters}\ }\textbf {\bibinfo {volume} {95}},\ \bibinfo {pages} {060501} (\bibinfo {year} {2005})}\BibitemShut {NoStop}%
\bibitem [{\citenamefont {Niemeyer}\ and\ \citenamefont {Kose}(1976)}]{Niemeyer1976}%
  \BibitemOpen
  \bibfield  {author} {\bibinfo {author} {\bibfnamefont {J.}~\bibnamefont {Niemeyer}}\ and\ \bibinfo {author} {\bibfnamefont {V.}~\bibnamefont {Kose}},\ }\bibfield  {title} {\bibinfo {title} {Observation of large dc supercurrents at nonzero voltages in josephson tunnel junctions},\ }\href {https://doi.org/10.1063/1.89094} {\bibfield  {journal} {\bibinfo  {journal} {Applied Physics Letters}\ }\textbf {\bibinfo {volume} {29}},\ \bibinfo {pages} {380} (\bibinfo {year} {1976})}\BibitemShut {NoStop}%
\bibitem [{\citenamefont {Dolan}(1977)}]{Dolan1977}%
  \BibitemOpen
  \bibfield  {author} {\bibinfo {author} {\bibfnamefont {G.~J.}\ \bibnamefont {Dolan}},\ }\bibfield  {title} {\bibinfo {title} {Offset masks for lift-off photoprocessing},\ }\href {https://doi.org/10.1063/1.89690} {\bibfield  {journal} {\bibinfo  {journal} {Applied Physics Letters}\ }\textbf {\bibinfo {volume} {31}},\ \bibinfo {pages} {337} (\bibinfo {year} {1977})}\BibitemShut {NoStop}%
\bibitem [{\citenamefont {Lecocq}\ \emph {et~al.}(2011)\citenamefont {Lecocq}, \citenamefont {Pop}, \citenamefont {Peng}, \citenamefont {Matei}, \citenamefont {Crozes}, \citenamefont {Fournier}, \citenamefont {Naud}, \citenamefont {Guichard},\ and\ \citenamefont {Buisson}}]{Lecocq2011}%
  \BibitemOpen
  \bibfield  {author} {\bibinfo {author} {\bibfnamefont {F.}~\bibnamefont {Lecocq}}, \bibinfo {author} {\bibfnamefont {I.~M.}\ \bibnamefont {Pop}}, \bibinfo {author} {\bibfnamefont {Z.}~\bibnamefont {Peng}}, \bibinfo {author} {\bibfnamefont {I.}~\bibnamefont {Matei}}, \bibinfo {author} {\bibfnamefont {T.}~\bibnamefont {Crozes}}, \bibinfo {author} {\bibfnamefont {T.}~\bibnamefont {Fournier}}, \bibinfo {author} {\bibfnamefont {C.}~\bibnamefont {Naud}}, \bibinfo {author} {\bibfnamefont {W.}~\bibnamefont {Guichard}},\ and\ \bibinfo {author} {\bibfnamefont {O.}~\bibnamefont {Buisson}},\ }\bibfield  {title} {\bibinfo {title} {Junction fabrication by shadow evaporation without a suspended bridge},\ }\href {https://doi.org/10.1088/0957-4484/22/31/315302} {\bibfield  {journal} {\bibinfo  {journal} {Nanotechnology}\ }\textbf {\bibinfo {volume} {22}},\ \bibinfo {pages} {315302} (\bibinfo {year} {2011})}\BibitemShut {NoStop}%
\bibitem [{\citenamefont {Ranadive}\ \emph {et~al.}(2022)\citenamefont {Ranadive}, \citenamefont {Esposito}, \citenamefont {Planat}, \citenamefont {Bonet}, \citenamefont {Naud}, \citenamefont {Buisson}, \citenamefont {Guichard},\ and\ \citenamefont {Roch}}]{Ranadive2022}%
  \BibitemOpen
  \bibfield  {author} {\bibinfo {author} {\bibfnamefont {A.}~\bibnamefont {Ranadive}}, \bibinfo {author} {\bibfnamefont {M.}~\bibnamefont {Esposito}}, \bibinfo {author} {\bibfnamefont {L.}~\bibnamefont {Planat}}, \bibinfo {author} {\bibfnamefont {E.}~\bibnamefont {Bonet}}, \bibinfo {author} {\bibfnamefont {C.}~\bibnamefont {Naud}}, \bibinfo {author} {\bibfnamefont {O.}~\bibnamefont {Buisson}}, \bibinfo {author} {\bibfnamefont {W.}~\bibnamefont {Guichard}},\ and\ \bibinfo {author} {\bibfnamefont {N.}~\bibnamefont {Roch}},\ }\bibfield  {title} {\bibinfo {title} {Kerr reversal in {Josephson} meta-material and traveling wave parametric amplification},\ }\href {https://doi.org/10.1038/s41467-022-29375-5} {\bibfield  {journal} {\bibinfo  {journal} {Nature Communications}\ }\textbf {\bibinfo {volume} {13}},\ \bibinfo {pages} {1737} (\bibinfo {year} {2022})}\BibitemShut {NoStop}%
\bibitem [{\citenamefont {Walter}\ \emph {et~al.}(2017)\citenamefont {Walter}, \citenamefont {Kurpiers}, \citenamefont {Gasparinetti}, \citenamefont {Magnard}, \citenamefont {Potočnik}, \citenamefont {Salathé}, \citenamefont {Pechal}, \citenamefont {Mondal}, \citenamefont {Oppliger}, \citenamefont {Eichler},\ and\ \citenamefont {Wallraff}}]{Walter2017}%
  \BibitemOpen
  \bibfield  {author} {\bibinfo {author} {\bibfnamefont {T.}~\bibnamefont {Walter}}, \bibinfo {author} {\bibfnamefont {P.}~\bibnamefont {Kurpiers}}, \bibinfo {author} {\bibfnamefont {S.}~\bibnamefont {Gasparinetti}}, \bibinfo {author} {\bibfnamefont {P.}~\bibnamefont {Magnard}}, \bibinfo {author} {\bibfnamefont {A.}~\bibnamefont {Potočnik}}, \bibinfo {author} {\bibfnamefont {Y.}~\bibnamefont {Salathé}}, \bibinfo {author} {\bibfnamefont {M.}~\bibnamefont {Pechal}}, \bibinfo {author} {\bibfnamefont {M.}~\bibnamefont {Mondal}}, \bibinfo {author} {\bibfnamefont {M.}~\bibnamefont {Oppliger}}, \bibinfo {author} {\bibfnamefont {C.}~\bibnamefont {Eichler}},\ and\ \bibinfo {author} {\bibfnamefont {A.}~\bibnamefont {Wallraff}},\ }\bibfield  {title} {\bibinfo {title} {Rapid {High}-{Fidelity} {Single}-{Shot} {Dispersive} {Readout} of {Superconducting} {Qubits}},\ }\href {https://doi.org/10.1103/PhysRevApplied.7.054020} {\bibfield  {journal} {\bibinfo  {journal} {Physical Review Applied}\ }\textbf {\bibinfo {volume}
  {7}},\ \bibinfo {pages} {054020} (\bibinfo {year} {2017})}\BibitemShut {NoStop}%
\bibitem [{\citenamefont {Spring}\ \emph {et~al.}(2024)\citenamefont {Spring}, \citenamefont {Milanovic}, \citenamefont {Sunada}, \citenamefont {Wang}, \citenamefont {Loo}, \citenamefont {Tamate},\ and\ \citenamefont {Nakamura}}]{Spring2024}%
  \BibitemOpen
  \bibfield  {author} {\bibinfo {author} {\bibfnamefont {P.}~\bibnamefont {Spring}}, \bibinfo {author} {\bibfnamefont {L.}~\bibnamefont {Milanovic}}, \bibinfo {author} {\bibfnamefont {Y.}~\bibnamefont {Sunada}}, \bibinfo {author} {\bibfnamefont {S.}~\bibnamefont {Wang}}, \bibinfo {author} {\bibfnamefont {A.}~\bibnamefont {Loo}}, \bibinfo {author} {\bibfnamefont {S.}~\bibnamefont {Tamate}},\ and\ \bibinfo {author} {\bibfnamefont {Y.}~\bibnamefont {Nakamura}},\ }\bibfield  {title} {\bibinfo {title} {Fast multiplexed superconducting qubit readout with intrinsic {Purcell} filtering}} (\bibinfo {year} {2024}),\ \bibinfo {note} {arXiv:2409.04967 [quant-ph]}\BibitemShut {NoStop}%
\bibitem [{\citenamefont {Krinner}\ \emph {et~al.}(2022)\citenamefont {Krinner}, \citenamefont {Lacroix}, \citenamefont {Remm}, \citenamefont {Di~Paolo}, \citenamefont {Genois}, \citenamefont {Leroux}, \citenamefont {Hellings}, \citenamefont {Lazar}, \citenamefont {Swiadek}, \citenamefont {Herrmann}, \citenamefont {Norris}, \citenamefont {Andersen}, \citenamefont {Müller}, \citenamefont {Blais}, \citenamefont {Eichler},\ and\ \citenamefont {Wallraff}}]{Krinner2022}%
  \BibitemOpen
  \bibfield  {author} {\bibinfo {author} {\bibfnamefont {S.}~\bibnamefont {Krinner}}, \bibinfo {author} {\bibfnamefont {N.}~\bibnamefont {Lacroix}}, \bibinfo {author} {\bibfnamefont {A.}~\bibnamefont {Remm}}, \bibinfo {author} {\bibfnamefont {A.}~\bibnamefont {Di~Paolo}}, \bibinfo {author} {\bibfnamefont {E.}~\bibnamefont {Genois}}, \bibinfo {author} {\bibfnamefont {C.}~\bibnamefont {Leroux}}, \bibinfo {author} {\bibfnamefont {C.}~\bibnamefont {Hellings}}, \bibinfo {author} {\bibfnamefont {S.}~\bibnamefont {Lazar}}, \bibinfo {author} {\bibfnamefont {F.}~\bibnamefont {Swiadek}}, \bibinfo {author} {\bibfnamefont {J.}~\bibnamefont {Herrmann}}, \bibinfo {author} {\bibfnamefont {G.~J.}\ \bibnamefont {Norris}}, \bibinfo {author} {\bibfnamefont {C.~K.}\ \bibnamefont {Andersen}}, \bibinfo {author} {\bibfnamefont {M.}~\bibnamefont {Müller}}, \bibinfo {author} {\bibfnamefont {A.}~\bibnamefont {Blais}}, \bibinfo {author} {\bibfnamefont {C.}~\bibnamefont {Eichler}},\ and\ \bibinfo {author} {\bibfnamefont
  {A.}~\bibnamefont {Wallraff}},\ }\bibfield  {title} {\bibinfo {title} {Realizing repeated quantum error correction in a distance-three surface code},\ }\href {https://doi.org/10.1038/s41586-022-04566-8} {\bibfield  {journal} {\bibinfo  {journal} {Nature}\ }\textbf {\bibinfo {volume} {605}},\ \bibinfo {pages} {669} (\bibinfo {year} {2022})}\BibitemShut {NoStop}%
\bibitem [{\citenamefont {Miao}\ \emph {et~al.}(2023)\citenamefont {Miao}, \citenamefont {McEwen}, \citenamefont {Atalaya}, \citenamefont {Kafri}, \citenamefont {Pryadko}, \citenamefont {Bengtsson}, \citenamefont {Opremcak}, \citenamefont {Satzinger}, \citenamefont {Chen}, \citenamefont {Klimov}, \citenamefont {Quintana}, \citenamefont {Acharya}, \citenamefont {Anderson}, \citenamefont {Ansmann}, \citenamefont {Arute}, \citenamefont {Arya}, \citenamefont {Asfaw}, \citenamefont {Bardin}, \citenamefont {Bourassa}, \citenamefont {Bovaird}, \citenamefont {Brill}, \citenamefont {Buckley}, \citenamefont {Buell}, \citenamefont {Burger}, \citenamefont {Burkett}, \citenamefont {Bushnell}, \citenamefont {Campero}, \citenamefont {Chiaro}, \citenamefont {Collins}, \citenamefont {Conner}, \citenamefont {Crook}, \citenamefont {Curtin}, \citenamefont {Debroy}, \citenamefont {Demura}, \citenamefont {Dunsworth}, \citenamefont {Erickson}, \citenamefont {Fatemi}, \citenamefont {Ferreira}, \citenamefont {Burgos}, \citenamefont
  {Forati}, \citenamefont {Fowler}, \citenamefont {Foxen}, \citenamefont {Garcia}, \citenamefont {Giang}, \citenamefont {Gidney}, \citenamefont {Giustina}, \citenamefont {Gosula}, \citenamefont {Dau}, \citenamefont {Gross}, \citenamefont {Hamilton}, \citenamefont {Harrington}, \citenamefont {Heu}, \citenamefont {Hilton}, \citenamefont {Hoffmann}, \citenamefont {Hong}, \citenamefont {Huang}, \citenamefont {Huff}, \citenamefont {Iveland}, \citenamefont {Jeffrey}, \citenamefont {Jiang}, \citenamefont {Jones}, \citenamefont {Kelly}, \citenamefont {Kim}, \citenamefont {Kostritsa}, \citenamefont {Kreikebaum}, \citenamefont {Landhuis}, \citenamefont {Laptev}, \citenamefont {Laws}, \citenamefont {Lee}, \citenamefont {Lester}, \citenamefont {Lill}, \citenamefont {Liu}, \citenamefont {Locharla}, \citenamefont {Lucero}, \citenamefont {Martin}, \citenamefont {Megrant}, \citenamefont {Mi}, \citenamefont {Montazeri}, \citenamefont {Morvan}, \citenamefont {Naaman}, \citenamefont {Neeley}, \citenamefont {Neill},
  \citenamefont {Nersisyan}, \citenamefont {Newman}, \citenamefont {Ng}, \citenamefont {Nguyen}, \citenamefont {Nguyen}, \citenamefont {Potter}, \citenamefont {Rocque}, \citenamefont {Roushan}, \citenamefont {Sankaragomathi}, \citenamefont {Schurkus}, \citenamefont {Schuster}, \citenamefont {Shearn}, \citenamefont {Shorter}, \citenamefont {Shutty}, \citenamefont {Shvarts}, \citenamefont {Skruzny}, \citenamefont {Smith}, \citenamefont {Sterling}, \citenamefont {Szalay}, \citenamefont {Thor}, \citenamefont {Torres}, \citenamefont {White}, \citenamefont {Woo}, \citenamefont {Yao}, \citenamefont {Yeh}, \citenamefont {Yoo}, \citenamefont {Young}, \citenamefont {Zalcman}, \citenamefont {Zhu}, \citenamefont {Zobrist}, \citenamefont {Neven}, \citenamefont {Smelyanskiy}, \citenamefont {Petukhov}, \citenamefont {Korotkov}, \citenamefont {Sank},\ and\ \citenamefont {Chen}}]{Miao2023}%
  \BibitemOpen
  \bibfield  {author} {\bibinfo {author} {\bibfnamefont {K.~C.}\ \bibnamefont {Miao}}, \bibinfo {author} {\bibfnamefont {M.}~\bibnamefont {McEwen}}, \bibinfo {author} {\bibfnamefont {J.}~\bibnamefont {Atalaya}}, \bibinfo {author} {\bibfnamefont {D.}~\bibnamefont {Kafri}}, \bibinfo {author} {\bibfnamefont {L.~P.}\ \bibnamefont {Pryadko}}, \bibinfo {author} {\bibfnamefont {A.}~\bibnamefont {Bengtsson}}, \bibinfo {author} {\bibfnamefont {A.}~\bibnamefont {Opremcak}}, \bibinfo {author} {\bibfnamefont {K.~J.}\ \bibnamefont {Satzinger}}, \bibinfo {author} {\bibfnamefont {Z.}~\bibnamefont {Chen}}, \bibinfo {author} {\bibfnamefont {P.~V.}\ \bibnamefont {Klimov}}, \bibinfo {author} {\bibfnamefont {C.}~\bibnamefont {Quintana}}, \bibinfo {author} {\bibfnamefont {R.}~\bibnamefont {Acharya}}, \bibinfo {author} {\bibfnamefont {K.}~\bibnamefont {Anderson}}, \bibinfo {author} {\bibfnamefont {M.}~\bibnamefont {Ansmann}}, \bibinfo {author} {\bibfnamefont {F.}~\bibnamefont {Arute}}, \bibinfo {author} {\bibfnamefont
  {K.}~\bibnamefont {Arya}}, \bibinfo {author} {\bibfnamefont {A.}~\bibnamefont {Asfaw}}, \bibinfo {author} {\bibfnamefont {J.~C.}\ \bibnamefont {Bardin}}, \bibinfo {author} {\bibfnamefont {A.}~\bibnamefont {Bourassa}}, \bibinfo {author} {\bibfnamefont {J.}~\bibnamefont {Bovaird}}, \bibinfo {author} {\bibfnamefont {L.}~\bibnamefont {Brill}}, \bibinfo {author} {\bibfnamefont {B.~B.}\ \bibnamefont {Buckley}}, \bibinfo {author} {\bibfnamefont {D.~A.}\ \bibnamefont {Buell}}, \bibinfo {author} {\bibfnamefont {T.}~\bibnamefont {Burger}}, \bibinfo {author} {\bibfnamefont {B.}~\bibnamefont {Burkett}}, \bibinfo {author} {\bibfnamefont {N.}~\bibnamefont {Bushnell}}, \bibinfo {author} {\bibfnamefont {J.}~\bibnamefont {Campero}}, \bibinfo {author} {\bibfnamefont {B.}~\bibnamefont {Chiaro}}, \bibinfo {author} {\bibfnamefont {R.}~\bibnamefont {Collins}}, \bibinfo {author} {\bibfnamefont {P.}~\bibnamefont {Conner}}, \bibinfo {author} {\bibfnamefont {A.~L.}\ \bibnamefont {Crook}}, \bibinfo {author} {\bibfnamefont
  {B.}~\bibnamefont {Curtin}}, \bibinfo {author} {\bibfnamefont {D.~M.}\ \bibnamefont {Debroy}}, \bibinfo {author} {\bibfnamefont {S.}~\bibnamefont {Demura}}, \bibinfo {author} {\bibfnamefont {A.}~\bibnamefont {Dunsworth}}, \bibinfo {author} {\bibfnamefont {C.}~\bibnamefont {Erickson}}, \bibinfo {author} {\bibfnamefont {R.}~\bibnamefont {Fatemi}}, \bibinfo {author} {\bibfnamefont {V.~S.}\ \bibnamefont {Ferreira}}, \bibinfo {author} {\bibfnamefont {L.~F.}\ \bibnamefont {Burgos}}, \bibinfo {author} {\bibfnamefont {E.}~\bibnamefont {Forati}}, \bibinfo {author} {\bibfnamefont {A.~G.}\ \bibnamefont {Fowler}}, \bibinfo {author} {\bibfnamefont {B.}~\bibnamefont {Foxen}}, \bibinfo {author} {\bibfnamefont {G.}~\bibnamefont {Garcia}}, \bibinfo {author} {\bibfnamefont {W.}~\bibnamefont {Giang}}, \bibinfo {author} {\bibfnamefont {C.}~\bibnamefont {Gidney}}, \bibinfo {author} {\bibfnamefont {M.}~\bibnamefont {Giustina}}, \bibinfo {author} {\bibfnamefont {R.}~\bibnamefont {Gosula}}, \bibinfo {author} {\bibfnamefont
  {A.~G.}\ \bibnamefont {Dau}}, \bibinfo {author} {\bibfnamefont {J.~A.}\ \bibnamefont {Gross}}, \bibinfo {author} {\bibfnamefont {M.~C.}\ \bibnamefont {Hamilton}}, \bibinfo {author} {\bibfnamefont {S.~D.}\ \bibnamefont {Harrington}}, \bibinfo {author} {\bibfnamefont {P.}~\bibnamefont {Heu}}, \bibinfo {author} {\bibfnamefont {J.}~\bibnamefont {Hilton}}, \bibinfo {author} {\bibfnamefont {M.~R.}\ \bibnamefont {Hoffmann}}, \bibinfo {author} {\bibfnamefont {S.}~\bibnamefont {Hong}}, \bibinfo {author} {\bibfnamefont {T.}~\bibnamefont {Huang}}, \bibinfo {author} {\bibfnamefont {A.}~\bibnamefont {Huff}}, \bibinfo {author} {\bibfnamefont {J.}~\bibnamefont {Iveland}}, \bibinfo {author} {\bibfnamefont {E.}~\bibnamefont {Jeffrey}}, \bibinfo {author} {\bibfnamefont {Z.}~\bibnamefont {Jiang}}, \bibinfo {author} {\bibfnamefont {C.}~\bibnamefont {Jones}}, \bibinfo {author} {\bibfnamefont {J.}~\bibnamefont {Kelly}}, \bibinfo {author} {\bibfnamefont {S.}~\bibnamefont {Kim}}, \bibinfo {author} {\bibfnamefont {F.}~\bibnamefont
  {Kostritsa}}, \bibinfo {author} {\bibfnamefont {J.~M.}\ \bibnamefont {Kreikebaum}}, \bibinfo {author} {\bibfnamefont {D.}~\bibnamefont {Landhuis}}, \bibinfo {author} {\bibfnamefont {P.}~\bibnamefont {Laptev}}, \bibinfo {author} {\bibfnamefont {L.}~\bibnamefont {Laws}}, \bibinfo {author} {\bibfnamefont {K.}~\bibnamefont {Lee}}, \bibinfo {author} {\bibfnamefont {B.~J.}\ \bibnamefont {Lester}}, \bibinfo {author} {\bibfnamefont {A.~T.}\ \bibnamefont {Lill}}, \bibinfo {author} {\bibfnamefont {W.}~\bibnamefont {Liu}}, \bibinfo {author} {\bibfnamefont {A.}~\bibnamefont {Locharla}}, \bibinfo {author} {\bibfnamefont {E.}~\bibnamefont {Lucero}}, \bibinfo {author} {\bibfnamefont {S.}~\bibnamefont {Martin}}, \bibinfo {author} {\bibfnamefont {A.}~\bibnamefont {Megrant}}, \bibinfo {author} {\bibfnamefont {X.}~\bibnamefont {Mi}}, \bibinfo {author} {\bibfnamefont {S.}~\bibnamefont {Montazeri}}, \bibinfo {author} {\bibfnamefont {A.}~\bibnamefont {Morvan}}, \bibinfo {author} {\bibfnamefont {O.}~\bibnamefont {Naaman}},
  \bibinfo {author} {\bibfnamefont {M.}~\bibnamefont {Neeley}}, \bibinfo {author} {\bibfnamefont {C.}~\bibnamefont {Neill}}, \bibinfo {author} {\bibfnamefont {A.}~\bibnamefont {Nersisyan}}, \bibinfo {author} {\bibfnamefont {M.}~\bibnamefont {Newman}}, \bibinfo {author} {\bibfnamefont {J.~H.}\ \bibnamefont {Ng}}, \bibinfo {author} {\bibfnamefont {A.}~\bibnamefont {Nguyen}}, \bibinfo {author} {\bibfnamefont {M.}~\bibnamefont {Nguyen}}, \bibinfo {author} {\bibfnamefont {R.}~\bibnamefont {Potter}}, \bibinfo {author} {\bibfnamefont {C.}~\bibnamefont {Rocque}}, \bibinfo {author} {\bibfnamefont {P.}~\bibnamefont {Roushan}}, \bibinfo {author} {\bibfnamefont {K.}~\bibnamefont {Sankaragomathi}}, \bibinfo {author} {\bibfnamefont {H.~F.}\ \bibnamefont {Schurkus}}, \bibinfo {author} {\bibfnamefont {C.}~\bibnamefont {Schuster}}, \bibinfo {author} {\bibfnamefont {M.~J.}\ \bibnamefont {Shearn}}, \bibinfo {author} {\bibfnamefont {A.}~\bibnamefont {Shorter}}, \bibinfo {author} {\bibfnamefont {N.}~\bibnamefont {Shutty}},
  \bibinfo {author} {\bibfnamefont {V.}~\bibnamefont {Shvarts}}, \bibinfo {author} {\bibfnamefont {J.}~\bibnamefont {Skruzny}}, \bibinfo {author} {\bibfnamefont {W.~C.}\ \bibnamefont {Smith}}, \bibinfo {author} {\bibfnamefont {G.}~\bibnamefont {Sterling}}, \bibinfo {author} {\bibfnamefont {M.}~\bibnamefont {Szalay}}, \bibinfo {author} {\bibfnamefont {D.}~\bibnamefont {Thor}}, \bibinfo {author} {\bibfnamefont {A.}~\bibnamefont {Torres}}, \bibinfo {author} {\bibfnamefont {T.}~\bibnamefont {White}}, \bibinfo {author} {\bibfnamefont {B.~W.~K.}\ \bibnamefont {Woo}}, \bibinfo {author} {\bibfnamefont {Z.~J.}\ \bibnamefont {Yao}}, \bibinfo {author} {\bibfnamefont {P.}~\bibnamefont {Yeh}}, \bibinfo {author} {\bibfnamefont {J.}~\bibnamefont {Yoo}}, \bibinfo {author} {\bibfnamefont {G.}~\bibnamefont {Young}}, \bibinfo {author} {\bibfnamefont {A.}~\bibnamefont {Zalcman}}, \bibinfo {author} {\bibfnamefont {N.}~\bibnamefont {Zhu}}, \bibinfo {author} {\bibfnamefont {N.}~\bibnamefont {Zobrist}}, \bibinfo {author}
  {\bibfnamefont {H.}~\bibnamefont {Neven}}, \bibinfo {author} {\bibfnamefont {V.}~\bibnamefont {Smelyanskiy}}, \bibinfo {author} {\bibfnamefont {A.}~\bibnamefont {Petukhov}}, \bibinfo {author} {\bibfnamefont {A.~N.}\ \bibnamefont {Korotkov}}, \bibinfo {author} {\bibfnamefont {D.}~\bibnamefont {Sank}},\ and\ \bibinfo {author} {\bibfnamefont {Y.}~\bibnamefont {Chen}},\ }\bibfield  {title} {\bibinfo {title} {Overcoming leakage in quantum error correction},\ }\href {https://doi.org/10.1038/s41567-023-02226-w} {\bibfield  {journal} {\bibinfo  {journal} {Nature Physics}\ }\textbf {\bibinfo {volume} {19}},\ \bibinfo {pages} {1780} (\bibinfo {year} {2023})}\BibitemShut {NoStop}%
\bibitem [{\citenamefont {Pereira}\ \emph {et~al.}(2022)\citenamefont {Pereira}, \citenamefont {García-Ripoll},\ and\ \citenamefont {Ramos}}]{Pereira2022}%
  \BibitemOpen
  \bibfield  {author} {\bibinfo {author} {\bibfnamefont {L.}~\bibnamefont {Pereira}}, \bibinfo {author} {\bibfnamefont {J.}~\bibnamefont {García-Ripoll}},\ and\ \bibinfo {author} {\bibfnamefont {T.}~\bibnamefont {Ramos}},\ }\bibfield  {title} {\bibinfo {title} {Complete {Physical} {Characterization} of {Quantum} {Nondemolition} {Measurements} via {Tomography}},\ }\href {https://doi.org/10.1103/PhysRevLett.129.010402} {\bibfield  {journal} {\bibinfo  {journal} {Physical Review Letters}\ }\textbf {\bibinfo {volume} {129}},\ \bibinfo {pages} {010402} (\bibinfo {year} {2022})}\BibitemShut {NoStop}%
\bibitem [{\citenamefont {Pereira}\ \emph {et~al.}(2023)\citenamefont {Pereira}, \citenamefont {García-Ripoll},\ and\ \citenamefont {Ramos}}]{Pereira2023}%
  \BibitemOpen
  \bibfield  {author} {\bibinfo {author} {\bibfnamefont {L.}~\bibnamefont {Pereira}}, \bibinfo {author} {\bibfnamefont {J.~J.}\ \bibnamefont {García-Ripoll}},\ and\ \bibinfo {author} {\bibfnamefont {T.}~\bibnamefont {Ramos}},\ }\bibfield  {title} {\bibinfo {title} {Parallel tomography of quantum non-demolition measurements in multi-qubit devices},\ }\href {https://doi.org/10.1038/s41534-023-00688-7} {\bibfield  {journal} {\bibinfo  {journal} {npj Quantum Information}\ }\textbf {\bibinfo {volume} {9}},\ \bibinfo {pages} {22} (\bibinfo {year} {2023})}\BibitemShut {NoStop}%
\bibitem [{\citenamefont {Hazra}\ \emph {et~al.}(2024)\citenamefont {Hazra}, \citenamefont {Dai}, \citenamefont {Connolly}, \citenamefont {Kurilovich}, \citenamefont {Wang}, \citenamefont {Frunzio},\ and\ \citenamefont {Devoret}}]{Hazra2024}%
  \BibitemOpen
  \bibfield  {author} {\bibinfo {author} {\bibfnamefont {S.}~\bibnamefont {Hazra}}, \bibinfo {author} {\bibfnamefont {W.}~\bibnamefont {Dai}}, \bibinfo {author} {\bibfnamefont {T.}~\bibnamefont {Connolly}}, \bibinfo {author} {\bibfnamefont {P.}~\bibnamefont {Kurilovich}}, \bibinfo {author} {\bibfnamefont {Z.}~\bibnamefont {Wang}}, \bibinfo {author} {\bibfnamefont {L.}~\bibnamefont {Frunzio}},\ and\ \bibinfo {author} {\bibfnamefont {M.}~\bibnamefont {Devoret}},\ }\bibfield  {title} {\bibinfo {title} {Benchmarking the readout of a superconducting qubit for repeated measurements}} (\bibinfo {year} {2024}),\ \bibinfo {note} {arXiv:2407.10934 [quant-ph]}\BibitemShut {NoStop}%
\bibitem [{\citenamefont {Touzard}\ \emph {et~al.}(2019)\citenamefont {Touzard}, \citenamefont {Kou}, \citenamefont {Frattini}, \citenamefont {Sivak}, \citenamefont {Puri}, \citenamefont {Grimm}, \citenamefont {Frunzio}, \citenamefont {Shankar},\ and\ \citenamefont {Devoret}}]{Touzard2019}%
  \BibitemOpen
  \bibfield  {author} {\bibinfo {author} {\bibfnamefont {S.}~\bibnamefont {Touzard}}, \bibinfo {author} {\bibfnamefont {A.}~\bibnamefont {Kou}}, \bibinfo {author} {\bibfnamefont {N.}~\bibnamefont {Frattini}}, \bibinfo {author} {\bibfnamefont {V.}~\bibnamefont {Sivak}}, \bibinfo {author} {\bibfnamefont {S.}~\bibnamefont {Puri}}, \bibinfo {author} {\bibfnamefont {A.}~\bibnamefont {Grimm}}, \bibinfo {author} {\bibfnamefont {L.}~\bibnamefont {Frunzio}}, \bibinfo {author} {\bibfnamefont {S.}~\bibnamefont {Shankar}},\ and\ \bibinfo {author} {\bibfnamefont {M.}~\bibnamefont {Devoret}},\ }\bibfield  {title} {\bibinfo {title} {Gated {Conditional} {Displacement} {Readout} of {Superconducting} {Qubits}},\ }\href {https://doi.org/10.1103/PhysRevLett.122.080502} {\bibfield  {journal} {\bibinfo  {journal} {Physical Review Letters}\ }\textbf {\bibinfo {volume} {122}},\ \bibinfo {pages} {080502} (\bibinfo {year} {2019})}\BibitemShut {NoStop}%
\bibitem [{\citenamefont {Gambetta}\ \emph {et~al.}(2008)\citenamefont {Gambetta}, \citenamefont {Blais}, \citenamefont {Boissonneault}, \citenamefont {Houck}, \citenamefont {Schuster},\ and\ \citenamefont {Girvin}}]{Gambetta2008}%
  \BibitemOpen
  \bibfield  {author} {\bibinfo {author} {\bibfnamefont {J.}~\bibnamefont {Gambetta}}, \bibinfo {author} {\bibfnamefont {A.}~\bibnamefont {Blais}}, \bibinfo {author} {\bibfnamefont {M.}~\bibnamefont {Boissonneault}}, \bibinfo {author} {\bibfnamefont {A.~A.}\ \bibnamefont {Houck}}, \bibinfo {author} {\bibfnamefont {D.~I.}\ \bibnamefont {Schuster}},\ and\ \bibinfo {author} {\bibfnamefont {S.~M.}\ \bibnamefont {Girvin}},\ }\bibfield  {title} {\bibinfo {title} {Quantum trajectory approach to circuit {QED}: {Quantum} jumps and the {Zeno} effect},\ }\href {https://doi.org/10.1103/PhysRevA.77.012112} {\bibfield  {journal} {\bibinfo  {journal} {Physical Review A}\ }\textbf {\bibinfo {volume} {77}},\ \bibinfo {pages} {012112} (\bibinfo {year} {2008})}\BibitemShut {NoStop}%
\bibitem [{\citenamefont {Blais}\ \emph {et~al.}(2021)\citenamefont {Blais}, \citenamefont {Grimsmo}, \citenamefont {Girvin},\ and\ \citenamefont {Wallraff}}]{Blais2021}%
  \BibitemOpen
  \bibfield  {author} {\bibinfo {author} {\bibfnamefont {A.}~\bibnamefont {Blais}}, \bibinfo {author} {\bibfnamefont {A.~L.}\ \bibnamefont {Grimsmo}}, \bibinfo {author} {\bibfnamefont {S.}~\bibnamefont {Girvin}},\ and\ \bibinfo {author} {\bibfnamefont {A.}~\bibnamefont {Wallraff}},\ }\bibfield  {title} {\bibinfo {title} {Circuit quantum electrodynamics},\ }\href {https://doi.org/10.1103/RevModPhys.93.025005} {\bibfield  {journal} {\bibinfo  {journal} {Reviews of Modern Physics}\ }\textbf {\bibinfo {volume} {93}},\ \bibinfo {pages} {025005} (\bibinfo {year} {2021})}\BibitemShut {NoStop}%
\bibitem [{\citenamefont {Sank}\ \emph {et~al.}(2016)\citenamefont {Sank}, \citenamefont {Chen}, \citenamefont {Khezri}, \citenamefont {Kelly}, \citenamefont {Barends}, \citenamefont {Campbell}, \citenamefont {Chen}, \citenamefont {Chiaro}, \citenamefont {Dunsworth}, \citenamefont {Fowler}, \citenamefont {Jeffrey}, \citenamefont {Lucero}, \citenamefont {Megrant}, \citenamefont {Mutus}, \citenamefont {Neeley}, \citenamefont {Neill}, \citenamefont {O’Malley}, \citenamefont {Quintana}, \citenamefont {Roushan}, \citenamefont {Vainsencher}, \citenamefont {White}, \citenamefont {Wenner}, \citenamefont {Korotkov},\ and\ \citenamefont {Martinis}}]{Sank2016}%
  \BibitemOpen
  \bibfield  {author} {\bibinfo {author} {\bibfnamefont {D.}~\bibnamefont {Sank}}, \bibinfo {author} {\bibfnamefont {Z.}~\bibnamefont {Chen}}, \bibinfo {author} {\bibfnamefont {M.}~\bibnamefont {Khezri}}, \bibinfo {author} {\bibfnamefont {J.}~\bibnamefont {Kelly}}, \bibinfo {author} {\bibfnamefont {R.}~\bibnamefont {Barends}}, \bibinfo {author} {\bibfnamefont {B.}~\bibnamefont {Campbell}}, \bibinfo {author} {\bibfnamefont {Y.}~\bibnamefont {Chen}}, \bibinfo {author} {\bibfnamefont {B.}~\bibnamefont {Chiaro}}, \bibinfo {author} {\bibfnamefont {A.}~\bibnamefont {Dunsworth}}, \bibinfo {author} {\bibfnamefont {A.}~\bibnamefont {Fowler}}, \bibinfo {author} {\bibfnamefont {E.}~\bibnamefont {Jeffrey}}, \bibinfo {author} {\bibfnamefont {E.}~\bibnamefont {Lucero}}, \bibinfo {author} {\bibfnamefont {A.}~\bibnamefont {Megrant}}, \bibinfo {author} {\bibfnamefont {J.}~\bibnamefont {Mutus}}, \bibinfo {author} {\bibfnamefont {M.}~\bibnamefont {Neeley}}, \bibinfo {author} {\bibfnamefont {C.}~\bibnamefont {Neill}}, \bibinfo
  {author} {\bibfnamefont {P.}~\bibnamefont {O’Malley}}, \bibinfo {author} {\bibfnamefont {C.}~\bibnamefont {Quintana}}, \bibinfo {author} {\bibfnamefont {P.}~\bibnamefont {Roushan}}, \bibinfo {author} {\bibfnamefont {A.}~\bibnamefont {Vainsencher}}, \bibinfo {author} {\bibfnamefont {T.}~\bibnamefont {White}}, \bibinfo {author} {\bibfnamefont {J.}~\bibnamefont {Wenner}}, \bibinfo {author} {\bibfnamefont {A.~N.}\ \bibnamefont {Korotkov}},\ and\ \bibinfo {author} {\bibfnamefont {J.~M.}\ \bibnamefont {Martinis}},\ }\bibfield  {title} {\bibinfo {title} {Measurement-{Induced} {State} {Transitions} in a {Superconducting} {Qubit}: {Beyond} the {Rotating} {Wave} {Approximation}},\ }\href {https://doi.org/10.1103/PhysRevLett.117.190503} {\bibfield  {journal} {\bibinfo  {journal} {Physical Review Letters}\ }\textbf {\bibinfo {volume} {117}},\ \bibinfo {pages} {190503} (\bibinfo {year} {2016})}\BibitemShut {NoStop}%
\bibitem [{\citenamefont {Ong}\ \emph {et~al.}(2011)\citenamefont {Ong}, \citenamefont {Boissonneault}, \citenamefont {Mallet}, \citenamefont {Palacios-Laloy}, \citenamefont {Dewes}, \citenamefont {Doherty}, \citenamefont {Blais}, \citenamefont {Bertet}, \citenamefont {Vion},\ and\ \citenamefont {Esteve}}]{Ong2011}%
  \BibitemOpen
  \bibfield  {author} {\bibinfo {author} {\bibfnamefont {F.~R.}\ \bibnamefont {Ong}}, \bibinfo {author} {\bibfnamefont {M.}~\bibnamefont {Boissonneault}}, \bibinfo {author} {\bibfnamefont {F.}~\bibnamefont {Mallet}}, \bibinfo {author} {\bibfnamefont {A.}~\bibnamefont {Palacios-Laloy}}, \bibinfo {author} {\bibfnamefont {A.}~\bibnamefont {Dewes}}, \bibinfo {author} {\bibfnamefont {A.~C.}\ \bibnamefont {Doherty}}, \bibinfo {author} {\bibfnamefont {A.}~\bibnamefont {Blais}}, \bibinfo {author} {\bibfnamefont {P.}~\bibnamefont {Bertet}}, \bibinfo {author} {\bibfnamefont {D.}~\bibnamefont {Vion}},\ and\ \bibinfo {author} {\bibfnamefont {D.}~\bibnamefont {Esteve}},\ }\bibfield  {title} {\bibinfo {title} {Circuit {QED} with a {Nonlinear} {Resonator}: ac-{Stark} {Shift} and {Dephasing}},\ }\href {https://doi.org/10.1103/PhysRevLett.106.167002} {\bibfield  {journal} {\bibinfo  {journal} {Physical Review Letters}\ }\textbf {\bibinfo {volume} {106}},\ \bibinfo {pages} {167002} (\bibinfo {year} {2011})}\BibitemShut
  {NoStop}%
\bibitem [{\citenamefont {Eichler}\ and\ \citenamefont {Wallraff}(2014)}]{Eichler2014}%
  \BibitemOpen
  \bibfield  {author} {\bibinfo {author} {\bibfnamefont {C.}~\bibnamefont {Eichler}}\ and\ \bibinfo {author} {\bibfnamefont {A.}~\bibnamefont {Wallraff}},\ }\bibfield  {title} {\bibinfo {title} {Controlling the dynamic range of a {Josephson} parametric amplifier},\ }\href {https://doi.org/10.1140/epjqt2} {\bibfield  {journal} {\bibinfo  {journal} {EPJ Quantum Technology}\ }\textbf {\bibinfo {volume} {1}},\ \bibinfo {pages} {1} (\bibinfo {year} {2014})}\BibitemShut {NoStop}%
\bibitem [{\citenamefont {Eichler}\ \emph {et~al.}(2014)\citenamefont {Eichler}, \citenamefont {Salathe}, \citenamefont {Mlynek}, \citenamefont {Schmidt},\ and\ \citenamefont {Wallraff}}]{Eichler2014a}%
  \BibitemOpen
  \bibfield  {author} {\bibinfo {author} {\bibfnamefont {C.}~\bibnamefont {Eichler}}, \bibinfo {author} {\bibfnamefont {Y.}~\bibnamefont {Salathe}}, \bibinfo {author} {\bibfnamefont {J.}~\bibnamefont {Mlynek}}, \bibinfo {author} {\bibfnamefont {S.}~\bibnamefont {Schmidt}},\ and\ \bibinfo {author} {\bibfnamefont {A.}~\bibnamefont {Wallraff}},\ }\bibfield  {title} {\bibinfo {title} {Quantum-{Limited} {Amplification} and {Entanglement} in {Coupled} {Nonlinear} {Resonators}},\ }\href {https://doi.org/10.1103/PhysRevLett.113.110502} {\bibfield  {journal} {\bibinfo  {journal} {Physical Review Letters}\ }\textbf {\bibinfo {volume} {113}},\ \bibinfo {pages} {110502} (\bibinfo {year} {2014})}\BibitemShut {NoStop}%
\bibitem [{\citenamefont {Frattini}\ \emph {et~al.}(2018)\citenamefont {Frattini}, \citenamefont {Sivak}, \citenamefont {Lingenfelter}, \citenamefont {Shankar},\ and\ \citenamefont {Devoret}}]{Frattini2018}%
  \BibitemOpen
  \bibfield  {author} {\bibinfo {author} {\bibfnamefont {N.~E.}\ \bibnamefont {Frattini}}, \bibinfo {author} {\bibfnamefont {V.~V.}\ \bibnamefont {Sivak}}, \bibinfo {author} {\bibfnamefont {A.}~\bibnamefont {Lingenfelter}}, \bibinfo {author} {\bibfnamefont {S.}~\bibnamefont {Shankar}},\ and\ \bibinfo {author} {\bibfnamefont {M.~H.}\ \bibnamefont {Devoret}},\ }\bibfield  {title} {\bibinfo {title} {Optimizing the {Nonlinearity} and {Dissipation} of a {SNAIL} {Parametric} {Amplifier} for {Dynamic} {Range}},\ }\href {https://doi.org/10.1103/PhysRevApplied.10.054020} {\bibfield  {journal} {\bibinfo  {journal} {Physical Review Applied}\ }\textbf {\bibinfo {volume} {10}},\ \bibinfo {pages} {054020} (\bibinfo {year} {2018})}\BibitemShut {NoStop}%
\bibitem [{\citenamefont {Planat}\ \emph {et~al.}(2019)\citenamefont {Planat}, \citenamefont {Dassonneville}, \citenamefont {Martínez}, \citenamefont {Foroughi}, \citenamefont {Buisson}, \citenamefont {Hasch-Guichard}, \citenamefont {Naud}, \citenamefont {Vijay}, \citenamefont {Murch},\ and\ \citenamefont {Roch}}]{Planat2019}%
  \BibitemOpen
  \bibfield  {author} {\bibinfo {author} {\bibfnamefont {L.}~\bibnamefont {Planat}}, \bibinfo {author} {\bibfnamefont {R.}~\bibnamefont {Dassonneville}}, \bibinfo {author} {\bibfnamefont {J.~P.}\ \bibnamefont {Martínez}}, \bibinfo {author} {\bibfnamefont {F.}~\bibnamefont {Foroughi}}, \bibinfo {author} {\bibfnamefont {O.}~\bibnamefont {Buisson}}, \bibinfo {author} {\bibfnamefont {W.}~\bibnamefont {Hasch-Guichard}}, \bibinfo {author} {\bibfnamefont {C.}~\bibnamefont {Naud}}, \bibinfo {author} {\bibfnamefont {R.}~\bibnamefont {Vijay}}, \bibinfo {author} {\bibfnamefont {K.}~\bibnamefont {Murch}},\ and\ \bibinfo {author} {\bibfnamefont {N.}~\bibnamefont {Roch}},\ }\bibfield  {title} {\bibinfo {title} {Understanding the {Saturation} {Power} of {Josephson} {Parametric} {Amplifiers} {Made} from {SQUID} {Arrays}},\ }\href {https://doi.org/10.1103/PhysRevApplied.11.034014} {\bibfield  {journal} {\bibinfo  {journal} {Physical Review Applied}\ }\textbf {\bibinfo {volume} {11}},\ \bibinfo {pages} {034014} (\bibinfo
  {year} {2019})}\BibitemShut {NoStop}%
\bibitem [{\citenamefont {Planat}\ \emph {et~al.}(2020)\citenamefont {Planat}, \citenamefont {Ranadive}, \citenamefont {Dassonneville}, \citenamefont {Puertas~Martínez}, \citenamefont {Léger}, \citenamefont {Naud}, \citenamefont {Buisson}, \citenamefont {Hasch-Guichard}, \citenamefont {Basko},\ and\ \citenamefont {Roch}}]{Planat2020}%
  \BibitemOpen
  \bibfield  {author} {\bibinfo {author} {\bibfnamefont {L.}~\bibnamefont {Planat}}, \bibinfo {author} {\bibfnamefont {A.}~\bibnamefont {Ranadive}}, \bibinfo {author} {\bibfnamefont {R.}~\bibnamefont {Dassonneville}}, \bibinfo {author} {\bibfnamefont {J.}~\bibnamefont {Puertas~Martínez}}, \bibinfo {author} {\bibfnamefont {S.}~\bibnamefont {Léger}}, \bibinfo {author} {\bibfnamefont {C.}~\bibnamefont {Naud}}, \bibinfo {author} {\bibfnamefont {O.}~\bibnamefont {Buisson}}, \bibinfo {author} {\bibfnamefont {W.}~\bibnamefont {Hasch-Guichard}}, \bibinfo {author} {\bibfnamefont {D.~M.}\ \bibnamefont {Basko}},\ and\ \bibinfo {author} {\bibfnamefont {N.}~\bibnamefont {Roch}},\ }\bibfield  {title} {\bibinfo {title} {Photonic-{Crystal} {Josephson} {Traveling}-{Wave} {Parametric} {Amplifier}},\ }\href {https://doi.org/10.1103/PhysRevX.10.021021} {\bibfield  {journal} {\bibinfo  {journal} {Physical Review X}\ }\textbf {\bibinfo {volume} {10}},\ \bibinfo {pages} {021021} (\bibinfo {year} {2020})}\BibitemShut {NoStop}%
\bibitem [{\citenamefont {Hu}\ \emph {et~al.}(2004)\citenamefont {Hu}, \citenamefont {Sarveswaran}, \citenamefont {Lieberman},\ and\ \citenamefont {Bernstein}}]{Hu2004}%
  \BibitemOpen
  \bibfield  {author} {\bibinfo {author} {\bibfnamefont {W.}~\bibnamefont {Hu}}, \bibinfo {author} {\bibfnamefont {K.}~\bibnamefont {Sarveswaran}}, \bibinfo {author} {\bibfnamefont {M.}~\bibnamefont {Lieberman}},\ and\ \bibinfo {author} {\bibfnamefont {G.~H.}\ \bibnamefont {Bernstein}},\ }\bibfield  {title} {\bibinfo {title} {Sub-10 nm electron beam lithography using cold development of poly(methylmethacrylate)},\ }\href {https://doi.org/10.1116/1.1763897} {\bibfield  {journal} {\bibinfo  {journal} {Journal of Vacuum Science \& Technology B: Microelectronics and Nanometer Structures Processing, Measurement, and Phenomena}\ }\textbf {\bibinfo {volume} {22}},\ \bibinfo {pages} {1711} (\bibinfo {year} {2004})}\BibitemShut {NoStop}%
\bibitem [{\citenamefont {Cord}\ \emph {et~al.}(2007)\citenamefont {Cord}, \citenamefont {Lutkenhaus},\ and\ \citenamefont {Berggren}}]{Cord2007}%
  \BibitemOpen
  \bibfield  {author} {\bibinfo {author} {\bibfnamefont {B.}~\bibnamefont {Cord}}, \bibinfo {author} {\bibfnamefont {J.}~\bibnamefont {Lutkenhaus}},\ and\ \bibinfo {author} {\bibfnamefont {K.~K.}\ \bibnamefont {Berggren}},\ }\bibfield  {title} {\bibinfo {title} {Optimal temperature for development of poly(methylmethacrylate)},\ }\href {https://doi.org/10.1116/1.2799978} {\bibfield  {journal} {\bibinfo  {journal} {Journal of Vacuum Science \& Technology B: Microelectronics and Nanometer Structures Processing, Measurement, and Phenomena}\ }\textbf {\bibinfo {volume} {25}},\ \bibinfo {pages} {2013} (\bibinfo {year} {2007})}\BibitemShut {NoStop}%
\bibitem [{\citenamefont {Ocola}\ and\ \citenamefont {Stein}(2006)}]{Ocola2006}%
  \BibitemOpen
  \bibfield  {author} {\bibinfo {author} {\bibfnamefont {L.~E.}\ \bibnamefont {Ocola}}\ and\ \bibinfo {author} {\bibfnamefont {A.}~\bibnamefont {Stein}},\ }\bibfield  {title} {\bibinfo {title} {Effect of cold development on improvement in electron-beam nanopatterning resolution and line roughness},\ }\href {https://doi.org/10.1116/1.2366698} {\bibfield  {journal} {\bibinfo  {journal} {Journal of Vacuum Science \& Technology B: Microelectronics and Nanometer Structures Processing, Measurement, and Phenomena}\ }\textbf {\bibinfo {volume} {24}},\ \bibinfo {pages} {3061} (\bibinfo {year} {2006})}\BibitemShut {NoStop}%
\bibitem [{\citenamefont {Braumüller}\ \emph {et~al.}(2016)\citenamefont {Braumüller}, \citenamefont {Sandberg}, \citenamefont {Vissers}, \citenamefont {Schneider}, \citenamefont {Schlör}, \citenamefont {Grünhaupt}, \citenamefont {Rotzinger}, \citenamefont {Marthaler}, \citenamefont {Lukashenko}, \citenamefont {Dieter}, \citenamefont {Ustinov}, \citenamefont {Weides},\ and\ \citenamefont {Pappas}}]{Braumueller2016}%
  \BibitemOpen
  \bibfield  {author} {\bibinfo {author} {\bibfnamefont {J.}~\bibnamefont {Braumüller}}, \bibinfo {author} {\bibfnamefont {M.}~\bibnamefont {Sandberg}}, \bibinfo {author} {\bibfnamefont {M.~R.}\ \bibnamefont {Vissers}}, \bibinfo {author} {\bibfnamefont {A.}~\bibnamefont {Schneider}}, \bibinfo {author} {\bibfnamefont {S.}~\bibnamefont {Schlör}}, \bibinfo {author} {\bibfnamefont {L.}~\bibnamefont {Grünhaupt}}, \bibinfo {author} {\bibfnamefont {H.}~\bibnamefont {Rotzinger}}, \bibinfo {author} {\bibfnamefont {M.}~\bibnamefont {Marthaler}}, \bibinfo {author} {\bibfnamefont {A.}~\bibnamefont {Lukashenko}}, \bibinfo {author} {\bibfnamefont {A.}~\bibnamefont {Dieter}}, \bibinfo {author} {\bibfnamefont {A.~V.}\ \bibnamefont {Ustinov}}, \bibinfo {author} {\bibfnamefont {M.}~\bibnamefont {Weides}},\ and\ \bibinfo {author} {\bibfnamefont {D.~P.}\ \bibnamefont {Pappas}},\ }\bibfield  {title} {\bibinfo {title} {Concentric transmon qubit featuring fast tunability and an anisotropic magnetic dipole moment},\ }\bibfield
  {journal} {\bibinfo  {journal} {Applied Physics Letters}\ }\textbf {\bibinfo {volume} {108}},\ \href {https://doi.org/10.1063/1.4940230} {10.1063/1.4940230} (\bibinfo {year} {2016})\BibitemShut {NoStop}%
\bibitem [{\citenamefont {Rahamim}\ \emph {et~al.}(2017)\citenamefont {Rahamim}, \citenamefont {Behrle}, \citenamefont {Peterer}, \citenamefont {Patterson}, \citenamefont {Spring}, \citenamefont {Tsunoda}, \citenamefont {Manenti}, \citenamefont {Tancredi},\ and\ \citenamefont {Leek}}]{Rahamim2017}%
  \BibitemOpen
  \bibfield  {author} {\bibinfo {author} {\bibfnamefont {J.}~\bibnamefont {Rahamim}}, \bibinfo {author} {\bibfnamefont {T.}~\bibnamefont {Behrle}}, \bibinfo {author} {\bibfnamefont {M.~J.}\ \bibnamefont {Peterer}}, \bibinfo {author} {\bibfnamefont {A.}~\bibnamefont {Patterson}}, \bibinfo {author} {\bibfnamefont {P.~A.}\ \bibnamefont {Spring}}, \bibinfo {author} {\bibfnamefont {T.}~\bibnamefont {Tsunoda}}, \bibinfo {author} {\bibfnamefont {R.}~\bibnamefont {Manenti}}, \bibinfo {author} {\bibfnamefont {G.}~\bibnamefont {Tancredi}},\ and\ \bibinfo {author} {\bibfnamefont {P.~J.}\ \bibnamefont {Leek}},\ }\bibfield  {title} {\bibinfo {title} {Double-sided coaxial circuit qed with out-of-plane wiring},\ }\bibfield  {journal} {\bibinfo  {journal} {Applied Physics Letters}\ }\textbf {\bibinfo {volume} {110}},\ \href {https://doi.org/10.1063/1.4984299} {10.1063/1.4984299} (\bibinfo {year} {2017})\BibitemShut {NoStop}%
\bibitem [{\citenamefont {Schuster}\ \emph {et~al.}(2005)\citenamefont {Schuster}, \citenamefont {Wallraff}, \citenamefont {Blais}, \citenamefont {Frunzio}, \citenamefont {Huang}, \citenamefont {Majer}, \citenamefont {Girvin},\ and\ \citenamefont {Schoelkopf}}]{Schuster2005}%
  \BibitemOpen
  \bibfield  {author} {\bibinfo {author} {\bibfnamefont {D.~I.}\ \bibnamefont {Schuster}}, \bibinfo {author} {\bibfnamefont {A.}~\bibnamefont {Wallraff}}, \bibinfo {author} {\bibfnamefont {A.}~\bibnamefont {Blais}}, \bibinfo {author} {\bibfnamefont {L.}~\bibnamefont {Frunzio}}, \bibinfo {author} {\bibfnamefont {R.-S.}\ \bibnamefont {Huang}}, \bibinfo {author} {\bibfnamefont {J.}~\bibnamefont {Majer}}, \bibinfo {author} {\bibfnamefont {S.~M.}\ \bibnamefont {Girvin}},\ and\ \bibinfo {author} {\bibfnamefont {R.~J.}\ \bibnamefont {Schoelkopf}},\ }\bibfield  {title} {\bibinfo {title} {ac {Stark} {Shift} and {Dephasing} of a {Superconducting} {Qubit} {Strongly} {Coupled} to a {Cavity} {Field}},\ }\href {https://doi.org/10.1103/PhysRevLett.94.123602} {\bibfield  {journal} {\bibinfo  {journal} {Physical Review Letters}\ }\textbf {\bibinfo {volume} {94}},\ \bibinfo {pages} {123602} (\bibinfo {year} {2005})}\BibitemShut {NoStop}%
\bibitem [{\citenamefont {Nguyen}\ \emph {et~al.}(2022)\citenamefont {Nguyen}, \citenamefont {Koolstra}, \citenamefont {Kim}, \citenamefont {Morvan}, \citenamefont {Chistolini}, \citenamefont {Singh}, \citenamefont {Nesterov}, \citenamefont {Jünger}, \citenamefont {Chen}, \citenamefont {Pedramrazi}, \citenamefont {Mitchell}, \citenamefont {Kreikebaum}, \citenamefont {Puri}, \citenamefont {Santiago},\ and\ \citenamefont {Siddiqi}}]{Nguyen2022}%
  \BibitemOpen
  \bibfield  {author} {\bibinfo {author} {\bibfnamefont {L.~B.}\ \bibnamefont {Nguyen}}, \bibinfo {author} {\bibfnamefont {G.}~\bibnamefont {Koolstra}}, \bibinfo {author} {\bibfnamefont {Y.}~\bibnamefont {Kim}}, \bibinfo {author} {\bibfnamefont {A.}~\bibnamefont {Morvan}}, \bibinfo {author} {\bibfnamefont {T.}~\bibnamefont {Chistolini}}, \bibinfo {author} {\bibfnamefont {S.}~\bibnamefont {Singh}}, \bibinfo {author} {\bibfnamefont {K.~N.}\ \bibnamefont {Nesterov}}, \bibinfo {author} {\bibfnamefont {C.}~\bibnamefont {Jünger}}, \bibinfo {author} {\bibfnamefont {L.}~\bibnamefont {Chen}}, \bibinfo {author} {\bibfnamefont {Z.}~\bibnamefont {Pedramrazi}}, \bibinfo {author} {\bibfnamefont {B.~K.}\ \bibnamefont {Mitchell}}, \bibinfo {author} {\bibfnamefont {J.~M.}\ \bibnamefont {Kreikebaum}}, \bibinfo {author} {\bibfnamefont {S.}~\bibnamefont {Puri}}, \bibinfo {author} {\bibfnamefont {D.~I.}\ \bibnamefont {Santiago}},\ and\ \bibinfo {author} {\bibfnamefont {I.}~\bibnamefont {Siddiqi}},\ }\bibfield  {title} {\bibinfo
  {title} {Blueprint for a {High}-{Performance} {Fluxonium} {Quantum} {Processor}},\ }\href {https://doi.org/10.1103/PRXQuantum.3.037001} {\bibfield  {journal} {\bibinfo  {journal} {PRX Quantum}\ }\textbf {\bibinfo {volume} {3}},\ \bibinfo {pages} {037001} (\bibinfo {year} {2022})}\BibitemShut {NoStop}%
\bibitem [{\citenamefont {Bertet}\ \emph {et~al.}(2005)\citenamefont {Bertet}, \citenamefont {Chiorescu}, \citenamefont {Burkard}, \citenamefont {Semba}, \citenamefont {Harmans}, \citenamefont {DiVincenzo},\ and\ \citenamefont {Mooij}}]{Bertet2005}%
  \BibitemOpen
  \bibfield  {author} {\bibinfo {author} {\bibfnamefont {P.}~\bibnamefont {Bertet}}, \bibinfo {author} {\bibfnamefont {I.}~\bibnamefont {Chiorescu}}, \bibinfo {author} {\bibfnamefont {G.}~\bibnamefont {Burkard}}, \bibinfo {author} {\bibfnamefont {K.}~\bibnamefont {Semba}}, \bibinfo {author} {\bibfnamefont {C.~J. P.~M.}\ \bibnamefont {Harmans}}, \bibinfo {author} {\bibfnamefont {D.~P.}\ \bibnamefont {DiVincenzo}},\ and\ \bibinfo {author} {\bibfnamefont {J.~E.}\ \bibnamefont {Mooij}},\ }\bibfield  {title} {\bibinfo {title} {Dephasing of a {Superconducting} {Qubit} {Induced} by {Photon} {Noise}},\ }\href {https://doi.org/10.1103/PhysRevLett.95.257002} {\bibfield  {journal} {\bibinfo  {journal} {Physical Review Letters}\ }\textbf {\bibinfo {volume} {95}},\ \bibinfo {pages} {257002} (\bibinfo {year} {2005})}\BibitemShut {NoStop}%
\end{thebibliography}%

\end{document}